\documentclass[twocolumn]{aastex63}
\usepackage[]{units}
\usepackage[update,prepend]{epstopdf}
\usepackage{graphicx}
\usepackage{amssymb}
\usepackage{amsmath}
\usepackage{threeparttable}
\usepackage{hyperref}
\usepackage{color}

\usepackage{mathtools}

\usepackage{makecell}
\usepackage{bm}
\usepackage{CJK}

\newcommand{\kms} {\,km\,s$^{-1}$}
\newcommand{\masyr} {\,mas\,yr$^{-1}$}

\newcommand{\Msun}{\,M$_\odot$}

\newcommand{\Nwidebinary}{800}
\newcommand{\Nhalowb}{33}
\newcommand{\Nparent}{100,967}
\newcommand{\Ngoodkinematics}{99,161}
\newcommand{\Nwbgoodkinematics}{798}
\newcommand{\Nwbdisk}{763}
\newcommand{\Nwbhalo}{33}
\newcommand{\Nwbgoodcolor}{584}
\newcommand{\Nwbdbprp}{546}

\mathchardef\mhyphen="2D
\shorttitle{Halo and thick-disk wide binaries}
\shortauthors{Hwang et al.}

\begin{document}

\title{Wide binaries from the H3 survey: the thick disk and halo have similar wide binary fractions}

\correspondingauthor{Hsiang-Chih Hwang}
\email{hchwang@ias.edu}

\author[0000-0003-4250-4437]{Hsiang-Chih Hwang}
\affiliation{Institute for Advanced Study, Princeton, 1 Einstein Drive, NJ 08540, USA}
\affiliation{Department of Physics \& Astronomy, Johns Hopkins University, Baltimore, MD 21218, USA}

\author[0000-0001-5082-9536]{Yuan-Sen Ting}
\affiliation{Research School of Astronomy \& Astrophysics, Australian National University, Cotter Rd., Weston, ACT 2611, Australia}
\affiliation{Research School of Computer Science, Australian National University, Acton ACT 2601, Australia}

\author[0000-0002-1590-8551]{Charlie Conroy}
\affiliation{Center for Astrophysics $\mid$ Harvard \& Smithsonian, 60 Garden St, Cambridge, MA 02138, USA}

\author[0000-0001-6100-6869]{Nadia L. Zakamska}
\affiliation{Institute for Advanced Study, Princeton, 1 Einstein Drive, NJ 08540, USA}
\affiliation{Department of Physics \& Astronomy, Johns Hopkins University, Baltimore, MD 21218, USA}

\author[0000-0002-6871-1752]{Kareem El-Badry}
\affiliation{Center for Astrophysics $\mid$ Harvard \& Smithsonian, 60 Garden St, Cambridge, MA 02138, USA}

\author[0000-0002-1617-8917]{Phillip Cargile}
\affiliation{Center for Astrophysics $\mid$ Harvard \& Smithsonian, 60 Garden St, Cambridge, MA 02138, USA}

\author[0000-0002-5177-727X]{Dennis Zaritsky}
\affiliation{Steward Observatory, University of Arizona, 933 North Cherry Avenue, Tucson, AZ 85721-0065, USA}

\author[0000-0002-0572-8012]{Vedant Chandra}
\affiliation{Center for Astrophysics $\mid$ Harvard \& Smithsonian, 60 Garden St, Cambridge, MA 02138, USA}

\author[0000-0002-6800-5778]{Jiwon Jesse Han}
\affiliation{Center for Astrophysics $\mid$ Harvard \& Smithsonian, 60 Garden St, Cambridge, MA 02138, USA}

\author[0000-0003-2573-9832]{Joshua S. Speagle (\begin{CJK*}{UTF8}{gbsn}沈佳士\ignorespacesafterend\end{CJK*})}
\altaffiliation{Banting \& Dunlap Fellow}
\affiliation{Department of Statistical Sciences, University of Toronto, Toronto, ON M5S 3G3, Canada}
\affiliation{David A. Dunlap Department of Astronomy \& Astrophysics, University of Toronto, Toronto, ON M5S 3H4, Canada}
\affiliation{Dunlap Institute for Astronomy \& Astrophysics, University of Toronto, Toronto, ON M5S 3H4, Canada}

\author[0000-0002-7846-9787]{Ana Bonaca}
\affiliation{Observatories of the Carnegie Institution for Science, 813 Santa Barbara Street, Pasadena, CA 91101, USA}

\begin{abstract}
Due to the different environments in the Milky Way's disk and halo, comparing wide binaries in the disk and halo is key to understanding wide binary formation and evolution. By using {\it Gaia} Early Data Release 3, we search for resolved wide binary companions in the H3 survey, a spectroscopic survey that has compiled $\sim$150,000 spectra for thick-disk and halo stars to date. We identify \Nwidebinary\ high-confidence (a contamination rate of 4\%) wide binaries and two resolved triples, with binary separations mostly between $10^3$-$10^5$\,AU and a lowest [Fe/H] of $-2.7$. Based on their Galactic kinematics, \Nhalowb\ of them are halo wide binaries, and most of those are associated with the accreted {\it Gaia}-Sausage-Enceladus galaxy. The wide binary fraction in the thick disk decreases toward the low metallicity end, consistent with the previous findings for the thin disk. Our key finding is that the halo wide binary fraction is consistent with the thick-disk stars at a fixed [Fe/H]. There is no significant dependence of the wide binary fraction on the $\alpha$-captured abundance. Therefore, the wide binary fraction is mainly determined by the iron abundance, not their disk or halo origin nor the $\alpha$-captured abundance. Our results suggest that the formation environments play a major role for the wide binary fraction, instead of other processes like radial migration that only apply to disk stars.

\end{abstract}


\section{Introduction}

Since wide binaries are weakly bound, they are sensitive to their formation and evolution environments. Wide binary fractions reflect the density of star formation environments because dense environments like star clusters can easily disrupt wide binaries and thus reduce wide binary formation \citep{Scally1999,Parker2009,Geller2019,Deacon2020}. In the Milky Way disk, gravitational interactions with passing stars, molecular clouds, and the Galactic tide can disrupt wide binaries at separations $\gtrsim 10^4$\,AU, providing a unique probe for pc-scale gravitational fields \citep{Retterer1982,Bahcall1985, Weinberg1987,Jiang2010a}. In the halo where the interactions with passing stars and molecular clouds are less important, halo wide binaries provide a critical constraint on the nature of dark matter \citep{Chaname2004,Quinn2009a, Carr2020}. Now with an increasing number of disrupted accreted galaxies found in the Milky Way's stellar halo \citep[][see review by \citealt{Helmi2020}]{Belokurov2018, Helmi2018,Conroy2019,Li2019,Belokurov2020a,Naidu2020}, halo wide binaries are critical in constraining wide binary formation and the formation environments of these progenitor dwarf galaxies in the early Universe.

Our recent work shows that the wide binary fraction (defined at $10^3$-$10^4$\,AU) in the disk peaks at the solar metallicity and decreases both toward the low- and high-metallicity ends \citep{Hwang2021a}. We propose several hypotheses that may explain the non-monotonic metallicity dependence, including the different stellar density in the formation environments at different epochs \citep{Harris1994,Elmegreen1997,Kravtsov2005, Kruijssen2014,Ma2020}, the dynamical unfolding of compact triples \citep{Reipurth2012,Elliott2016}, and the radial migration of Galactic orbits \citep{Sellwood2002}. In the radial migration scenario, stars with supersolar and subsolar metallicities are formed at other Galactocentric radii and then migrated to the solar neighborhood. Then the lower wide binary fraction at supersolar and subsolar metallicities may be due to the different formation environments at different Galactocentric radii, or due to that the radial migration process (e.g. the enhanced interaction with molecular clouds) can efficiently disrupt wide binaries.

To constrain these hypotheses, comparing properties of binaries in the disk and the halo is particularly valuable because they probe a very different metallicity range, formation environments, stellar ages, and evolution environments. In particular, halo stars do not experience radial migration in the Milky Way disk, and thus halo wide binaries do not get disrupted by the radial migration process. Therefore, if radial migration plays an important role in the wide binary fraction, we would expect a higher wide binary fraction in the halo than in the disk. 

Despite their importance, halo wide binaries have been challenging to identify in the past. First, the density of halo stars is low compared to that of the disk in the solar neighborhood, comprising less than 0.2 per cent of local stars \citep{Helmi2008}. Therefore, assembling a large and clean halo star sample is not straightforward. Second, identifying a resolved companion as the wide binary companion was difficult before the {\it Gaia} era. Even though the {\it Gaia} mission provides the proper motions and parallaxes for billions of stars \citep{Gaia2016}, most of the wide binary samples are still limited within 1\,kpc \citep{El-Badry2018b,Hartman2020,El-Badry2021}. Furthermore, most of these wide binaries do not have radial velocities available from {\it Gaia}, making differentiating disk and halo stars challenging. Therefore, the number of robustly identified halo wide binaries remains insufficient to place a useful constraint on the halo wide binary fraction \citep{Hwang2021a}.

In this paper, we overcome these difficulties by having a customized wide binary search for stars in the H3 survey. The H3 survey is a large spectroscopic survey targeting halo and thick-disk stars \citep{Conroy2019a}. There are $\sim150,000$ H3 stars in our current analysis, and it is expected to have $\sim$300,000 stars at the end of the survey, with about 20\%\ of them being halo stars. The high-precision radial velocities measured from H3 are critical to differentiate disk and halo stars by their Galactic kinematics. Since most of the H3 stars are more than 1\,kpc away, we optimize the wide binary method for the H3 stars to maximize the resulting information. With the H3 wide binaries, we investigate the difference in the wide binary fractions of the halo and the disk to understand the formation and evolution of wide binaries.

The paper is structured as follows. Sec.~\ref{sec:sec2} describes the H3 and {\it Gaia} dataset and the general criteria for the sample selection. Sec.~\ref{sec:search} details the method of wide binary search and presents the results. Sec.~\ref{sec:properties} explores the properties of the wide binary candidates, and Sec.~\ref{sec:wbf-abundance} investigates the metallicity dependence of the wide binary fraction. We then discuss the implications in Sec.~\ref{sec:discussion} and conclude in Sec.~\ref{sec:conclusion}. Throughout the paper, we use `wide binaries' and `wide binary candidates' interchangeably when referring to the selected wide binaries. We use `metallicity' and `iron abundance' ([Fe/H]) interchangeably and always refer to [$\alpha$/Fe] as $\alpha$-captured abundances. We use the notation `binary' even if some of them can be unresolved triples.

\section{Sample selection}
\label{sec:sec2}

\subsection{H3 survey}
\label{sec:h3}

The H3 (`Hectochelle in the Halo at High Resolution') survey is a high-resolution spectroscopic ($R\approx32,000$) survey targeting thick-disk and halo stars \citep{Conroy2019a}. The survey is conducted by the 6.5-m MMT with a wavelength coverage of 5150-5300\AA. The survey has been collecting data since 2017. The main selection of H3 is composed of the following criteria: (1) $15<r<18$ where $r$ is the $r-$band magnitude from Pan-STARRS \citep{Chambers2016,Flewelling2016}; (2) $\pi-2\sigma_{\pi}<0.5$\,mas (later changed to $\pi<0.4$) where $\pi$ and $\sigma_{\pi}$ are parallaxes and parallax uncertainties from {\it Gaia} Data Release 2 \citep{Gaia2018Brown}; (3) $|b|>30^{\circ}$ to increase the halo star fraction; (4) declination $>-20^{\circ}$ to be observable from MMT. The H3 survey has secondary selections targeting stellar streams, K giants, blue horizontal branch stars, and RR Lyrae. This simple selection of magnitudes and parallaxes provides an unbiased view of the distant Milky Way.

The stellar parameters are measured from the H3 spectra using \texttt{MINESweeper} \citep{Cargile2020}. \texttt{MINESweeper} is a Bayesian framework that incorporates the information from H3 spectra, broadband photometry, and priors like {\it Gaia} parallaxes to fit stellar models computed by \texttt{MIST} (v2.0) \citep[][Dotter et al. in preparation]{Dotter2016,Choi2016}. \cite{Cargile2020} show that the stellar parameters of benchmark stars and clusters measured by \texttt{MINESweeper} are in good agreement with literature values.

In this paper, we use the radial velocities, distances, surface iron abundances ([Fe/H]), surface $\alpha$-captured elemental abundances ([$\alpha$/Fe]), and extinction ($A_V$) measured from \texttt{MINESweeper}. The surface elemental abundances may differ from the initial abundances due to atomic diffusion and mixing. By including the model from \cite{Dotter2017}, \texttt{MINESweeper} derives the initial chemical abundances, denoted as [Fe/H]$_{\rm init}$ and [$\alpha$/Fe]$_{\rm init}$. The difference between [Fe/H] and [Fe/H]$_{\rm init}$ depends on the evolutionary stages, stellar ages, and initial metallicities, with the largest difference of $0.5$\,dex happening at the main-sequence turnoff of metal-poor stars \citep{Dotter2017}. In our sample, the median difference between [Fe/H] and [Fe/H]$_{\rm init}$ is 0.1\,dex, and we do not find significant changes in our results between them. We present our results using surface abundances for better comparison with the literature. The $\alpha$-captured elemental abundances from \texttt{MINESweeper} scale the $\alpha$-elements O, Ne, Mg, Si, S, Ca, and Ti, and the measurement is most sensitive to [Mg/Fe] due to H3's wavelength coverage of Mg I triplet \citep{Cargile2020}.

In this paper, we use the proprietary H3 data from \texttt{rcat\_V4.0.3.d20201031\_MSG.fits}, with data collected between September 2017 and October 2020. We require \texttt{FLAG==0} and signal-to-noise ratios (SNR) $>3$ to avoid unreliable stellar parameter measurements. Targets from the Sagittarius stream (\texttt{Sgr\_FLAG==1} from \citealt{Johnson2020}) and tiles specifically for streams (\texttt{tileID} starting with `tb') are excluded because stream members moving in similar directions may increase the chance-alignment probability. For stars that were observed by H3 multiple times, we use the entries with the highest SNR. These selections result in a parent sample of \Nparent\ unique H3 stars for wide binary search. In this sample, the median uncertainty is 0.05\,dex for [Fe/H], 0.05\,dex for [$\alpha$/Fe], 0.2\kms\ for radial velocities, and 6\%\ for distances.

\subsection{Gaia survey}
\label{sec:data-gaia}

{\it Gaia} is an all-sky survey that provides optical broad-band photometry and high-precision astrometric measurements \citep{Gaia2016}. {\it Gaia} Early Data Release 3 (EDR3) \citep{Gaia2020Brown} was released in December 2020, with parallaxes and proper motions available for 1.5 billion sources down to {\it Gaia} G-band magnitudes of $\sim21$\,mag. In addition, {\it Gaia} measures radial velocities for bright stars ($G\lesssim15$\,mag for EDR3, and may reach $G\sim16$\,mag by the end of mission), but most halo stars are fainter than this magnitude limit. This is one of the H3 survey's motivations to target for stars fainter than 15\,mag. We use {\it Gaia} EDR3 to search for the comoving companions around the H3 stars. Because H3 target selection uses {\it Gaia}'s parallaxes which are derived together with proper motions, all H3 targets have proper motions from {\it Gaia}. 

The goal of this paper is to search for H3 stars' resolved wide companions that have nearly identical {\it Gaia} proper motions as the H3 targets. We query a field star sample from {\it Gaia} EDR3 to conduct the wide binary search. For this field star sample, we require that their proper motions be available in {\it Gaia} EDR3 (\texttt{astrometric\_params\_solved}$==31$ or 95). We do not apply any criteria on magnitudes and astrometric quality indicators (e.g. \texttt{ruwe}) to maximize the chance of finding the companions. Our wide binary selection compares the proper motions for each pair, which implicitly requires good astrometric quality, and it is inevitable that we may miss some genuine wide binaries because some of their stars have worse astrometric quality.

{\it Gaia} EDR3's ability to resolve close pairs starts dropping significantly below 0.7\arcsec, and only about 50\% of pairs can be spatially resolved at 0.5\arcsec\ \citep{Gaia2020Fabricius}. We do not use {\it Gaia}'s BP-RP colors for our wide binary selection, but we use them when investigating the properties of the selected wide binaries. When BP-RP colors are used in the analysis, we require their \texttt{bp\_flux\_over\_error} and \texttt{rp\_flux\_over\_error} to be larger than 10. Unlike G-band photometry, BP and RP fluxes do not have de-blending treatment, and therefore their fluxes can be affected by nearby sources within about 2\arcsec\ \citep{Riello2021}. To ensure that BP-RP colors are not strongly affected by the nearby sources, we apply an additional criterion of \texttt{phot\_bp\_rp\_excess\_factor}$<1.4$ when BP-RP colors are used in the analysis.

\subsection{Calculations of Galactic kinematics}
\label{sec:cri-kin}

Radial velocities from the H3 survey are necessary for computing Galactic kinematics parameters and to differentiating disk and halo stars. We measure the parameters for Galactic kinematics, including the total energy ($E_{tot}$) and the angular momentum along the Galactic $z$-direction ($L_z$), using \texttt{Gala} v1.1 \citep{gala2020}. We use the Galactocentric frame from \texttt{Astropy} v4.0 \citep{Astropy2013, Astropy2018}: Sun's Galactocentric radius $R_0=8.122$\,kpc \citep{GravityCollaboration2018}, solar motion with respect to the local standard of rest [$V_{R, \odot}$, $V_{\phi, \odot}$, $V_{Z, \odot}$]=[$-12.9$, $245.6$, $7.78$]\kms \citep{Drimmel2018}, and Sun's current Galactic height $Z_\odot=20.8$\,pc \citep{Bennett2019}. The Milky Way potential \texttt{MilkyWayPotential} \citep{Bovy2015} is adopted. $E_{tot}$ and $L_z$ are used to separate halo and disk stars, and using different Milky Way potential models does not affect this selection significantly \citep{Naidu2020}. A right-handed coordinate is used, and thus a star on a prograde (retrograde) orbit has a negative (positive) $L_z$. The Galactic orbit is integrated for 25 Gyr with a time step of 1 Myr using the \citet{Dormand1978} integration scheme. The eccentricity of Galactic orbits is then derived from $e=(r_{apo}-r_{peri})/(r_{apo}+r_{peri})$ where $r_{apo}$ and $r_{peri}$ are the apocenter and pericenter of the Galactic orbit. 

The observational inputs for kinematic parameter calculations are the radial velocities and distances measured from H3 spectra using \texttt{MINESweeper}, and celestial coordinates and astrometric solutions measured from {\it Gaia} EDR3. Following \cite{Naidu2020}, when $E_{tot}$ and $L_z$ are used in the analysis, we require that (i) $|E_{tot}|/\sigma_{E_{tot}}>3$ $\land$ $|L_z|/\sigma_{L_z}>3$; or (ii) $\sigma_{E_{tot}}<0.1\times10^5$\,km$^2$\,s$^{-2}$ $\land$ $\sigma_{L_z}<0.5\times10^3$\,kpc\,km\,s$^{-1}$, where `$\land$' is the Boolean `and' operator. The first condition is a normal 3-$\sigma$ uncertainty cut, and the second condition is to ensure that we keep halo stars with small $L_z$ in the sample (otherwise they may be excluded by the 3-$\sigma$ uncertainty cut). Out of \Nparent\ H3 stars, \Ngoodkinematics\ (98 per cent) of them satisfy these kinematic criteria.

\section{Wide binary search}
\label{sec:search}

\begin{figure*}
	\centering
	\includegraphics[height=.24\linewidth]{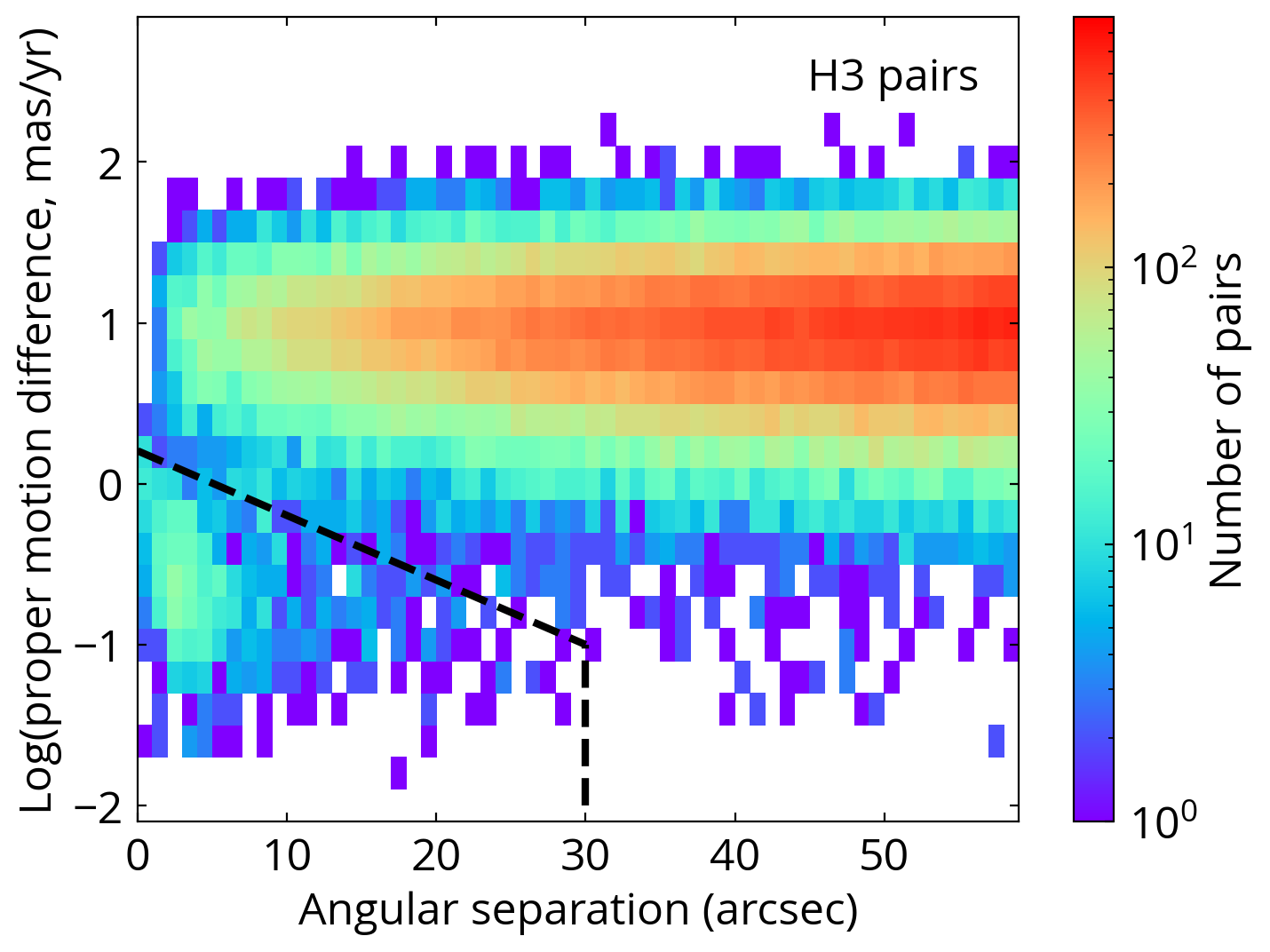}
	\includegraphics[height=.24\linewidth]{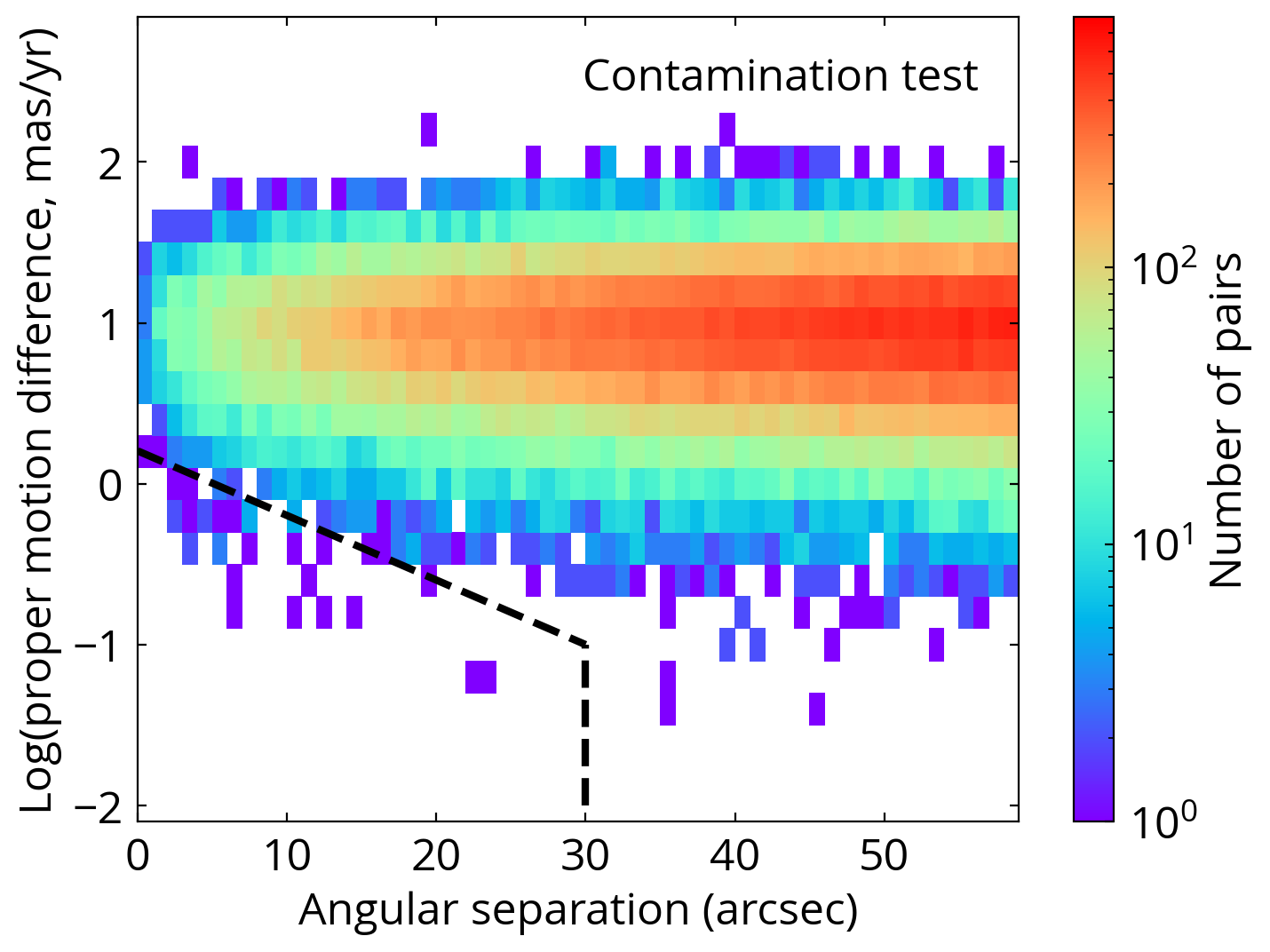}
	\includegraphics[height=.24\linewidth]{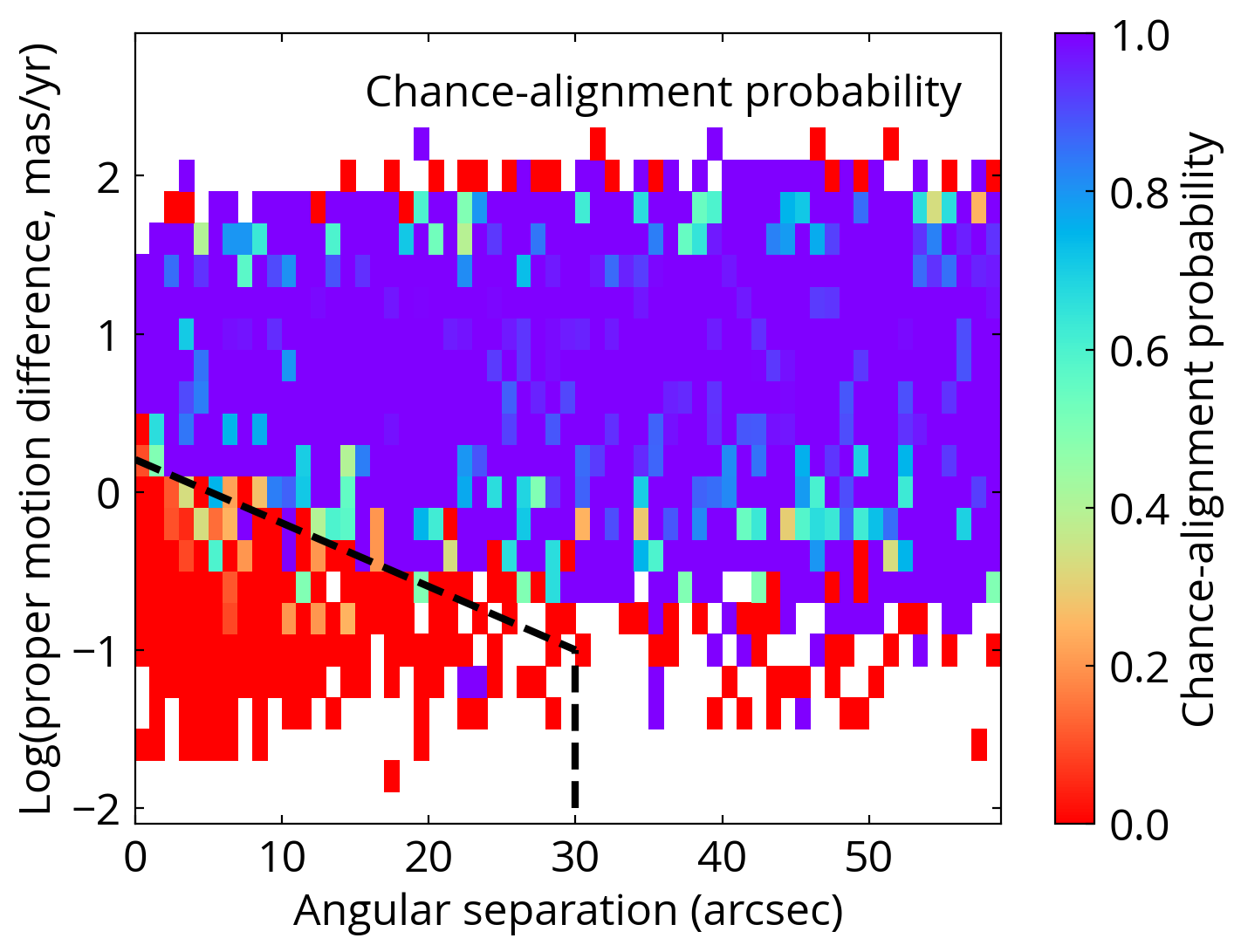}
	\caption{The wide binary search in the proper motion difference-angular separation space. Left: the distribution for all the pairs consisting of an H3 star and a {\it Gaia} field star. In addition to the chance-alignment pairs dominating in the upper-right part, there is an enhanced wide binary population with small proper motion differences and small angular separations. Middle: the contamination test where the H3 stars' Galactic latitudes are offset by 1 degree. Therefore, all the pairs in this panel are chance-alignment pairs. Right: the chance-alignment probability computed from the left and middle panel. Based on the chance-alignment probability, we use an empirical selection (the dashed lines in the three panels) to select high-confidence wide binaries. }
	\label{fig:WBsearch-one}
\end{figure*}

For a solar-mass wide binary with a semi-major axis of $>1000$\,AU, its orbital Keplerian velocity is $<0.7$\kms. Therefore, two resolved stars in a wide binary would have nearly identical proper motions. For a given H3 star and a nearby star from {\it Gaia}, we compute the proper motion difference by
\begin{equation}
\delta{PM} = \sqrt{({\rm \texttt{pmra}_1} - {\rm \texttt{pmra}_0})^2 + ({\rm \texttt{pmdec}_1} - {\rm \texttt{pmdec}_0})^2},
\end{equation}
where ${\rm \texttt{pmra}_i}$ and ${\rm \texttt{pmdec}_i}$ are the {\it Gaia} EDR3 proper motions in right ascension direction ($\mu_\alpha \cos \delta$) and in declination direction ($\mu_\delta$) for star $i$ ($=0$ or 1), respectively. With {\it Gaia}'s proper motion precision at the relevant magnitudes, the typical uncertainties of $\delta PM$ is $\sim 0.1$\masyr. The relative velocity difference projected in the plane of sky is $\delta V = 4.74 \times \delta PM \times d$, where $\delta V$ is in units of \kms, $\delta PM$ is in units of \masyr, $d$ is the heliocentric distance of the pair in kpc, and the factor of 4.74 is from the unit conversion. Therefore, for a wide binary at distances of 1 to 10\,kpc, {\it Gaia} can measure the difference in the projected orbital velocity $\delta V$ with uncertainties of $0.5-5$\kms. This precision is sufficient to distinguish genuine wide binaries where the orbital velocities are $<1$\kms\ from random chance-alignment pairs where the typical velocity differences are $>50$\kms\ for the thick-disk and halo stars.

For every H3 target, we search for a possible resolved wide companion in {\it Gaia} within 60\,arcsec. In most cases, radial velocities are only available for H3 targets but not for {\it Gaia} field stars, so we cannot compute radial velocity difference for the pair to help the binary search. Furthermore, limited by the fiber allocation, the minimum angular separation between two H3 stars is 20\,arcsec, making the wide binary search among H3 stars (when both stars were observed by H3) difficult to result in a statistically useful sample. Most H3 targets are distant ($>1$\,kpc) and therefore they do not have precise distances from {\it Gaia} EDR3. The precision of the distance measurement from {\it Gaia} and the \texttt{MINESweeper} fit is not sufficient to distinctly separate the wide binary population from the random chance-alignment pairs in the relative velocity-physical separation space (e.g. \citealt{Hwang2020c}). Without distances, proper motion differences and angular separations cannot be converted to projected relative velocity differences and physical separations. Therefore, we search the wide binaries in the proper motion difference-angular separation space. The advantage is that their measurement uncertainties are nearly negligible ($<0.1$\,mas for angular separations and $\sim0.1$\masyr\ for proper motion differences), with the caveat that the search has different sensitivities for binary's physical separations at different distances.

Fig.~\ref{fig:WBsearch-one} (left panel) shows the proper motion differences and angular separations of all the H3-{\it Gaia} pairs. To reduce the contamination, we require that the parallax difference of the pair is consistent with zero within $3\sigma_\pi$. An enhanced population with separations $<10$\,arcsec and proper motion differences $\lesssim1$\masyr can be seen, indicative of the resolved wide binary populations. The rest of the pairs in the upper-right part of the plot are chance alignments. 

To better quantify the contamination from the chance-alignment pairs, we conduct a test by offsetting the H3 targets' Galactic latitudes by 1 degree (1 degree corresponds to 17 pc at a distance of 1 kpc) to the North (with the coordinates of field {\it Gaia} stars unshifted) and redo the wide binary search \citep{Lepine2007}. Therefore, all the pairs from this test are chance-alignment pairs. The result is shown in Fig.~\ref{fig:WBsearch-one}, middle panel. This contamination test well reproduces the chance-alignment pairs in the left panel, except that the left panel has an additional wide binary population at small angular separations and small proper motion differences. 

In the right panel of Fig.~\ref{fig:WBsearch-one}, we compute the chance alignment probability (CAP) in each two-dimensional bin. Specifically, CAP in every bin is calculated by the number of pairs in the middle panel divided by the number of pairs in the left panel. The right panel of Fig.~\ref{fig:WBsearch-one} shows that the wide binaries can be selected by small proper motion differences with CAP less than 10\% out to 30\,arcsec. At separations $>30$\,arcsec, CAP is still less than $20$\% at small proper motion differences ($\lesssim0.3$\masyr), indicative of the existence of wide binaries beyond 30\,arcsec ($\sim0.1$\,pc for stars at 1\,kpc distances), consistent with the simulation and observation that a significant fraction of comoving stars with separations up to $\sim10$\,pc are conatal \citep{Kamdar2019,Nelson2021}. Some low CAP pixels at the edge of large proper motion differences are due to small-number statistics.

In this paper, we focus on the wide binaries with angular separations $<30$\,arcsec for their larger sample size and the lower CAP. We adopt an empirical demarcation line (the dashed lines in Fig.~\ref{fig:WBsearch-one}) to select high-confidence wide binaries with CAP$<10$\%. The demarcation line starts from $\delta{PM}=1.6$\masyr\, at 0\arcsec\ separation to $\delta{PM}=0.1$\masyr\ at 30\arcsec, connected by a straight line in the log-linear space. Pairs below this demarcation line and at angular separations $<30$\,arcsec are selected as wide binaries, resulting in \Nwidebinary\ wide binaries and 2 resolved triples. There are 33 pairs located below this demarcation line in the contamination test in the middle panel of  Fig.~\ref{fig:WBsearch-one}, suggesting that the contamination rate from the chance-alignment pairs in our wide binary sample is $4.1\pm0.7$ per cent. A more stringent selection (e.g. angular separations $<10$\,arcsec instead of $<30$\,arcsec) does not change our main results and conclusions.

We provide an electronic catalog for the H3 wide binaries. Table~\ref{tab:catalog} presents the data model for the catalog. Field names starting with the prefix `0\_' are the information for the H3 stars, and those starting with the prefix `1\_' are for the companions. A resolved triple system would have two entries in the table with the same H3 star and different companion stars.

\begin{table*}[]
	\centering
	\caption{Data model of the wide binary catalog. The full table is available as supplementary material. }
	\label{tab:catalog}
	\begin{tabular}{ll} 
		\hline \hline
		Field 							& Description \\
		\hline
		\texttt{0\_source\_id}    		  & {\it Gaia} EDR3 source\_id of the H3 star \\
		\texttt{0\_ra} 						    & Right ascension of the H3 star from {\it Gaia} EDR3 (J2016.0; deg) \\
		\texttt{0\_dec} 						   & Declination of the H3 star from {\it Gaia} EDR3 (J2016.0; deg) \\
		\texttt{0\_parallax} 				& Parallax of the H3 star from {\it Gaia} EDR3 (mas) \\
		\texttt{0\_parallax\_error}       & Uncertainty in \texttt{0\_parallax} from {\it Gaia} EDR3 (mas) \\
		\texttt{0\_pmra}		 				& Proper motion in right ascension direction of the H3 star from {\it Gaia} EDR3 (mas yr$^{-1}$) \\
		\texttt{0\_pmra\_error}           & Uncertainty in \texttt{0\_pmra} (mas yr$^{-1}$) from {\it Gaia} EDR3 \\
		\texttt{0\_pmdec}		 				& Proper motion in declination direction of the H3 star from {\it Gaia} EDR3 (mas yr$^{-1}$) \\
		\texttt{0\_pmdec\_error}           & Uncertainty in \texttt{0\_pmdec} (mas yr$^{-1}$) from {\it Gaia} EDR3 \\
		\texttt{0\_G}							& Apparent G-band magnitude of the H3 star from {\it Gaia} EDR3 (mag) \\
		\texttt{0\_H3\_ID}		&  H3 id for the H3 star \\ 
		\texttt{0\_FeH}                    & Iron abundance [Fe/H] of the H3 star measured by H3 (dex) \\
		\texttt{0\_aFe}                    & $\alpha$-captured elemental abundance [$\alpha$/Fe] of the H3 star measured by H3 (dex) \\
		\texttt{0\_Teff} 					   & Effective temperature of the H3 star measured by H3 (K) \\
		\texttt{0\_logg} 					  & Surface gravity of the H3 star measured by H3 ($\log$ cgs) \\
		\texttt{0\_Vrad}						& Radial velocity of the H3 star measured by H3 (\kms) \\
		\texttt{0\_dist\_adpt}            & Heliocentric distances of the H3 star measured by H3 (kpc) \\
		\texttt{0\_Etot}                     & Total orbital energy of the H3 star ($10^5$ km$^2$\,s$^{-2}$) \\
		\texttt{0\_Etot\_err}                     & Uncertainty in \texttt{0\_Etot} ($10^5$ km$^2$\,s$^{-2}$) \\
		\texttt{0\_Lz}                     & Orbital momentum along the Galactic z-axis of the H3 star ($10^3$ kpc\,km\,s$^{-1}$) \\
		\texttt{0\_Lz\_err}                     & Uncertainty in \texttt{0\_Lz} ($10^3$ kpc\,km\,s$^{-1}$) \\
		\texttt{1\_source\_id}    		  & {\it Gaia} EDR3 source\_id of the companion star \\
		\texttt{1\_ra} 						    & Right ascension of the companion star from {\it Gaia} EDR3 (J2016.0; deg) \\
		\texttt{1\_dec} 						   & Declination of the companion star from {\it Gaia} EDR3 (J2016.0; deg) \\
		\texttt{1\_parallax} 				& Parallax of the companion star from {\it Gaia} EDR3 (mas) \\
		\texttt{1\_parallax\_error}       & Uncertainty in \texttt{1\_parallax} from {\it Gaia} EDR3 (mas) \\
		\texttt{1\_pmra}		 				& Proper motion in right ascension direction of the companion star from {\it Gaia} EDR3 (mas yr$^{-1}$) \\
		\texttt{1\_pmra\_error}           & Uncertainty in \texttt{1\_pmra} (mas yr$^{-1}$) from {\it Gaia} EDR3 \\
		\texttt{1\_pmdec}		 				& Proper motion in declination direction of the companion star from {\it Gaia} EDR3 (mas yr$^{-1}$) \\
		\texttt{1\_pmdec\_error}           & Uncertainty in \texttt{1\_pmdec} (mas yr$^{-1}$) from {\it Gaia} EDR3 \\
		\texttt{1\_G}							& Apparent G-band magnitude of the companion star from {\it Gaia} EDR3 (mag) \\

		\texttt{separation\_arcsec}                  & Angular separation of the wide binary (arcsec) \\
		\texttt{separation\_AU}                  & Projected physical separation of the wide binary (AU) \\
		\texttt{pm\_diff}						& Proper motion difference of the wide binary (\masyr) \\
		\texttt{triple}  & 1 if it is a resolved triple, otherwise 0 \\
		\texttt{halo\_disk}  & 1 if it is in the halo; 0 if it is in the disk; $-1$ if it does not pass the kinematic criteria \\
		\hline \hline
		
	\end{tabular}
\end{table*}

\section{Properties of H3 wide binaries}

\label{sec:properties}

\subsection{Distances and binary separations}
\label{sec:dist-separation}

\begin{figure}
	\centering
	\includegraphics[width=1\linewidth]{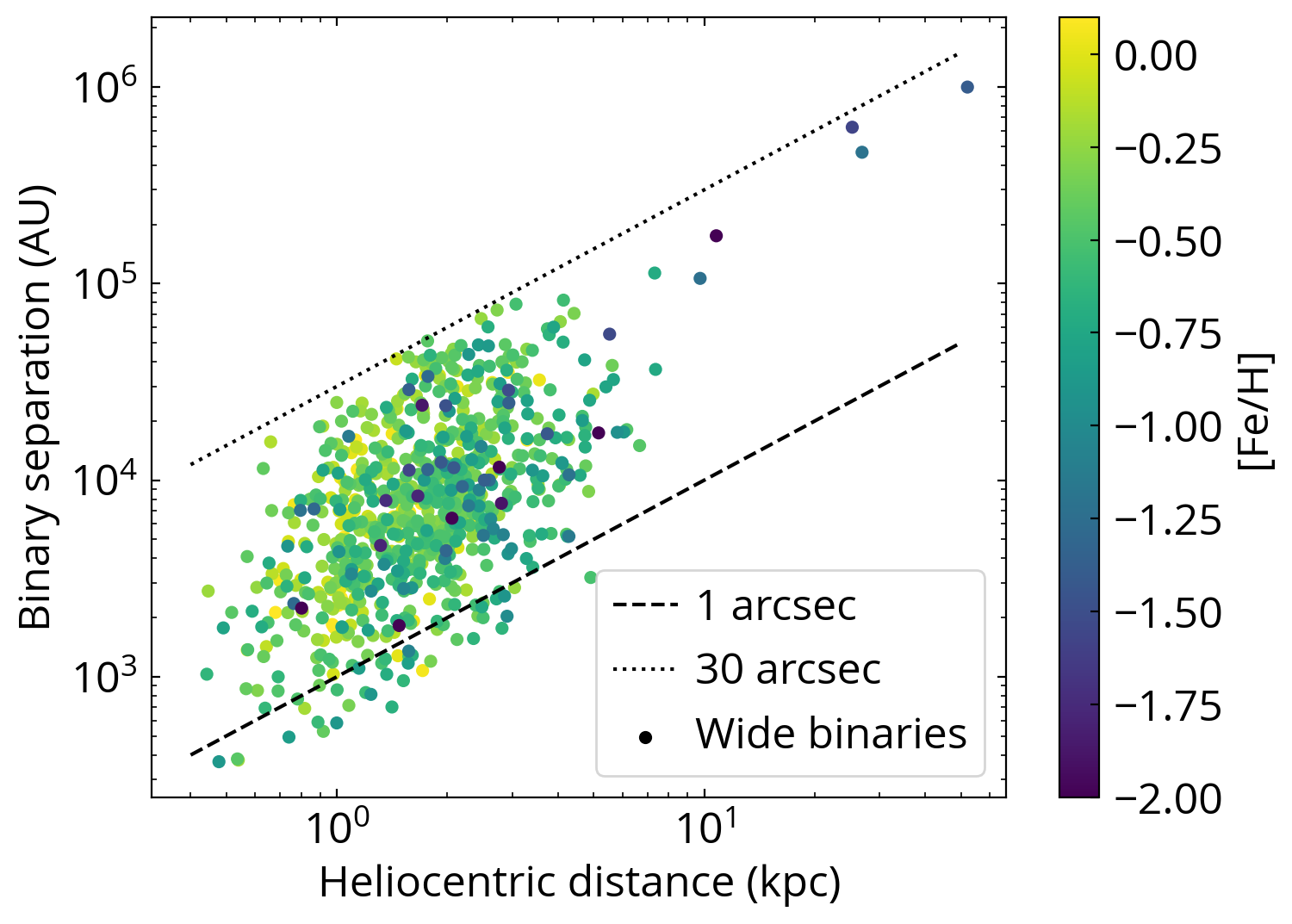}
	\caption{Binary separations and heliocentric distances of wide binaries, color-coded by their iron abundances. The dashed line corresponds to an angular separation of 1\,arcsec, roughly {\it Gaia}'s spatial resolution. The dotted line indicates 30\,arcsec used in our selection criterion. Most of the wide binaries are 1-5\,kpc from the Sun with binary separations of $10^3$-$10^5$\,AU. }
	\label{fig:sep-dist}
\end{figure}

Fig.~\ref{fig:sep-dist} shows the binary separations and heliocentric distances of the wide binaries, color-coded by their iron abundances. Most of the wide binaries have angular separations between 1 (dashed line) and 30 \,arcsec (dotted line), where the former is close to {\it Gaia}'s spatial resolution and the latter is due to our selection criterion. Most of these wide binaries are located 1-5\,kpc from the Sun. Those closer to the observer tend to have higher iron abundances.

There are four wide binaries located at larger than 10 kpc from the Sun, with the farthest one at 50\,kpc. They all have very wide binary separations of $\gtrsim1$\,pc. These distant wide binaries are valuable for probing the gravitational interactions in the outer Milky Way. Even though these distant wide binaries may suffer from a higher chance-alignment probability because of their larger angular separations ($>15$\,arcsec) and smaller total proper motions ($<5$\,\masyr), their companions' photometry, as shown in the next Section, are consistent with their H3 stars' isochrones, suggesting that they are not chance-alignment pairs. One possibility is that they are pairs with other member stars in a yet unidentified stream, instead of being a wide binary. This may explain why these four distant wide binaries all have large angular separations. Further work is needed to investigate if they are wide binaries or stream-member pairs.

\subsection{Hertzsprung-Russell diagram}
\label{sec:HR}

\begin{figure}
	\centering
	\includegraphics[width=1\linewidth]{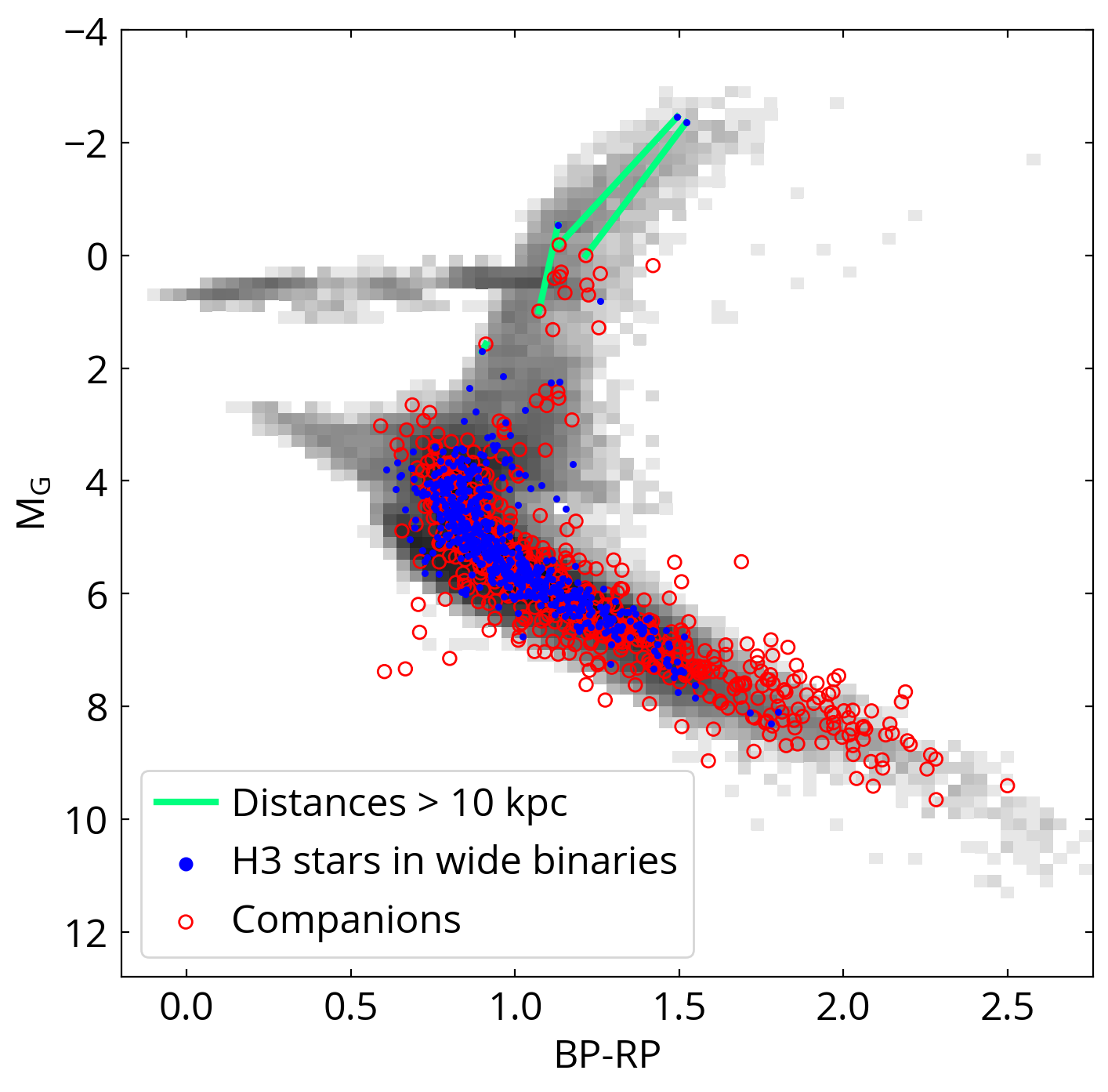}
	\caption{The H-R diagram of the wide binaries. The blue points show the H3 targets in the wide binaries, and the red open circles are their companions. Most of the H3 wide binaries are main-sequence stars. The sold green lines connect four wide binaries at distances $>10$\,kpc, and all of them are double-giant wide binaries, with one having nearly identical evolutionary stages for two member stars (giant twin). }
	\label{fig:HR}
\end{figure}

Fig.~\ref{fig:HR} shows the Hertzsprung-Russell (H-R) diagram of the wide binaries, where the absolute G-band magnitudes are computed using the distances measured from \texttt{MINESweeper}. The non-H3 stars adopt the same distances as their H3 companions. We apply the criteria on BP and RP as described in Sec.~\ref{sec:data-gaia} for more robust photometry, resulting in \Nwbgoodcolor\ pairs. The gray background shows the distribution of all stars in the H3 catalog, where most of them do not have wide companions. Most of the wide binaries identified here are main-sequence stars, and some are located on the giant branch. The companions scatter more around the main sequence track because they are fainter and more likely to have larger uncertainties in their photometry.

The solid green lines in Fig.~\ref{fig:HR} highlight the four wide binaries with distances $>10$\,kpc. They are rare double-giant wide binaries due to the selection effect that we only see giant stars at large distances. One of them has nearly the same G and BP-RP magnitudes for the two member stars, forming an unusual twin giant binary where two stars have nearly identical evolutionary stages. Fig.~\ref{fig:HR} shows that the non-H3 companions of these distant wide binaries have {\it Gaia} photometry consistent with the H3 stars' isochrones, suggesting that they are not chance-alignment pairs. 

Because there are limited regions in the H-R diagram where stars can exist, the locations of the non-H3 companions in the H-R diagram are a useful diagnostic of whether they are genuine binary companions. Since wide binaries with separations $\lesssim$ a few pc are conatal, two stars of the same wide binary have nearly identical stellar ages and chemical abundances \citep{Andrews2018,Kamdar2019,Hawkins2020,Nelson2021}. We use the distance, [Fe/H]$_{\rm init}$, and [$\alpha$/Fe]$_{\rm init}$ measured from the H3 stars to investigate whether the companion stars are located close to the corresponding \texttt{MIST} isochrone in the H-R diagram. We define color deviation $\Delta($BP-RP$)$ by:

\begin{equation}
\label{eq:color-deviation}
\Delta ({\rm BP}-{\rm RP})_{i} =  ({\rm BP}-{\rm RP})_{i} - ({\rm BP}-{\rm RP})_{iso, i},
\end{equation}
where ${\rm (BP-RP)}_{i}$ is the observed {\it Gaia} BP-RP color for star $i$ (a non-H3 companion), and ${\rm (BP-RP)}_{iso,i}$ is the expected BP-RP color from the MIST isochrone for this star given its absolute G-band magnitude. If the companion star $i$ is located close to the expected isochrone, then $\Delta {\rm (BP-RP)}_{i}\approx 0$. We use \texttt{brutus}\footnote{\url{https://github.com/joshspeagle/brutus}} (Speagle et al. submitted) to interpolate the MIST v2.0 grid \citep{Choi2016} given the [Fe/H]$_{\rm init}$, [$\alpha$/Fe]$_{\rm init}$, stellar age, and $A_V$ measured from the H3 star. Then for a non-H3 companion, its expected color ${\rm (BP-RP)}_{iso,i}$ is computed from the MIST isochrone given its absolute G-band magnitude, where we assume that the non-H3 companion has the same distance as the H3 star. This step also requires that an absolute G-band magnitude only maps to a single BP-RP value, which is satisfied for the main-sequence phase and the early stage of the giant phase. For the late giant phase, one absolute G-band magnitude can intersect an isochrone at multiple BP-RP colors. Therefore, we only compute $\Delta {\rm (BP-RP)}_{i}$ for stars where their ${\rm (BP-RP)}_{iso,i}$ can be uniquely determined, resulting in \Nwbdbprp\ pairs. Because \texttt{brutus} includes the extinction when computing ${\rm (BP-RP)}_{iso,i}$, the resulting $\Delta {\rm (BP-RP)}_{i}$ is extinction corrected.

\begin{figure}
	\centering
	\includegraphics[width=1\linewidth]{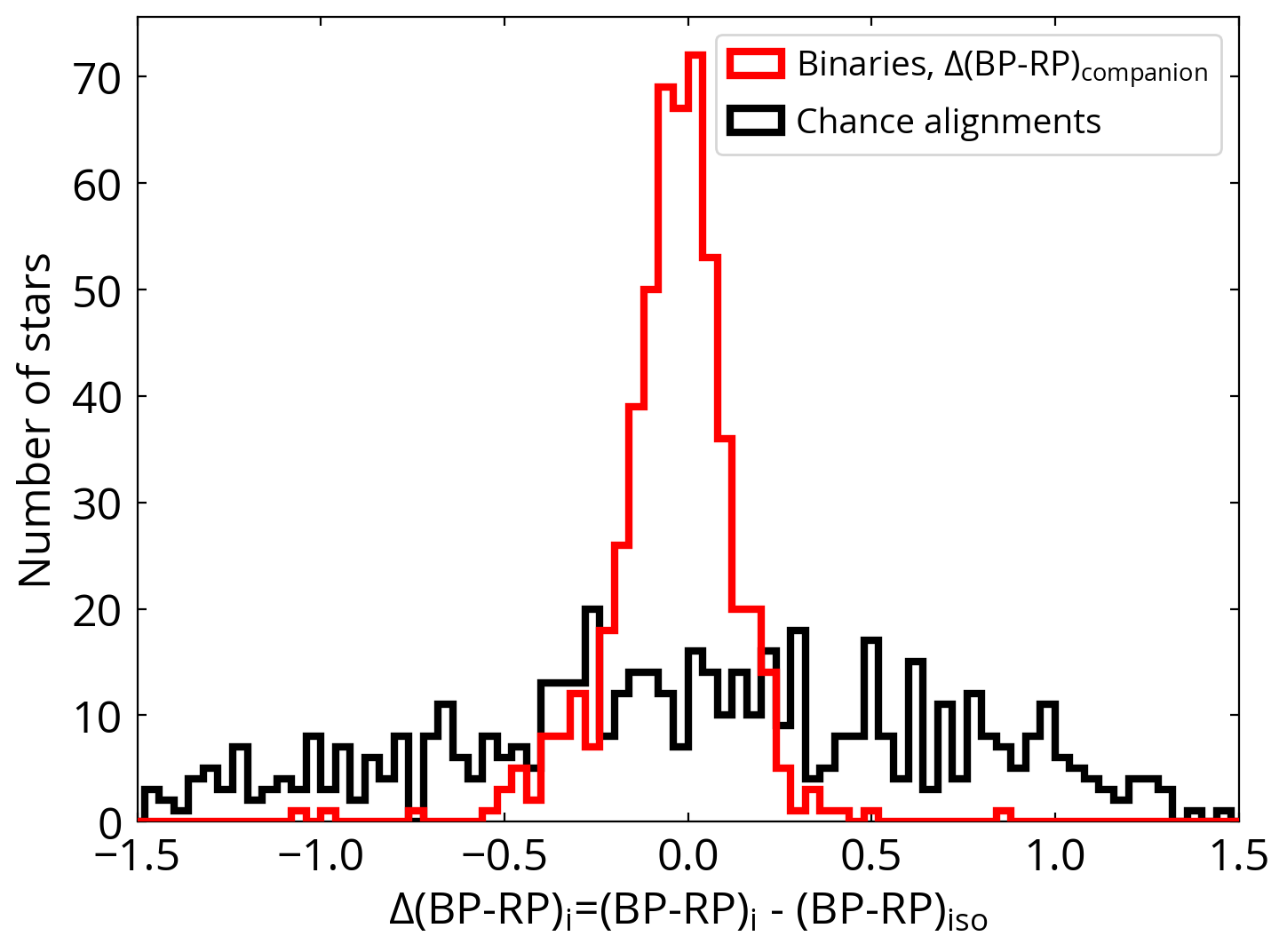}
	\caption{The distribution of color deviation $\Delta$(BP-RP), defined in Eq.~\ref{eq:color-deviation}. The color deviation of the companions around H3 stars (red) is strongly concentrated around 0. In contrast, the companions from chance-alignment pairs have a wide range of color deviations (black). This comparison suggests that most of the wide binaries are genuine and are not chance-alignment pairs. }
	\label{fig:dbprp}
\end{figure}

Fig.~\ref{fig:dbprp} shows the distribution of $\Delta {\rm (BP-RP)}$ for the non-H3 members of the wide binaries, $\Delta {\rm (BP-RP)}_{companion}$, and chance-alignment stars, $\Delta {\rm (BP-RP)}_{chance}$. The chance-alignment stars are selected by large proper motion differences $>5$\masyr\ and within 10\,arcsec around the H3 stars. The $\Delta {\rm (BP-RP)}_{companion}$ (red histogram) is clustered around 0, indicating that these companions are located close to the expected isochrone. The width of $\Delta {\rm (BP-RP)}_{companion}$ is mainly due to the uncertainties of distance estimates and photometry. For comparison, chance-alignment companions (black histogram) have a broad distribution of $\Delta {\rm (BP-RP)}_{chance}$. This is because these chance-alignment stars in general have different distances as the H3 stars, resulting in non-zero $\Delta {\rm (BP-RP)}_{chance}$ due to their erroneous absolute G-band magnitudes when we assume they have the same distances as the H3 stars. Therefore, Fig.~\ref{fig:dbprp} suggests that the majority of the wide binary candidates are indeed genuine wide binaries instead of chance-alignment stars.

\subsection{Galactic kinematics}
\label{sec:kinematics}

\begin{figure}
	\centering
	\includegraphics[width=1\linewidth]{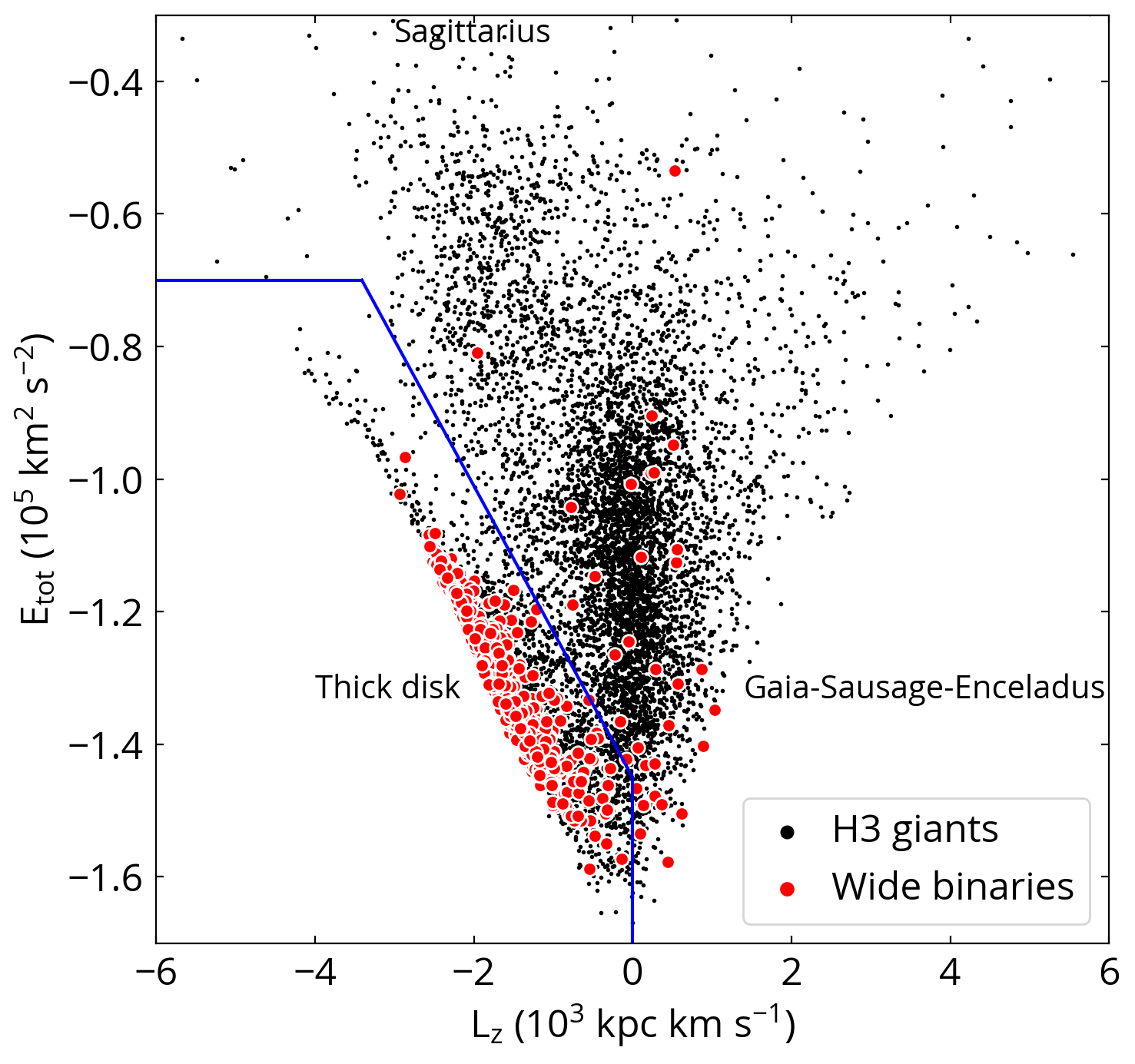}
	\caption{The Galactic orbits of stars in the $E_{tot}$-$L_z$ space. The H3 giant stars (black points) represent the disk and the kinematic substructures in the halo, with labels highlighting the prominent structures. H3 wide binaries are shown as red points. The empirical demarcation line used to distinguish the disk and halo stars is in blue. Most of the wide binaries are in the disk. There are 33 halo wide binaries, and most of them are associated with the accreted {\it Gaia}-Sausage-Enceladus remnant, which dominates the halo stars in the inner Milky Way. }
	\label{fig:E-Lz}
\end{figure}

Galactic kinematics is key to revealing substructures in the Milky Way's stellar halo \citep{Johnston1996,Helmi1999,Bullock2005,Cooper2010a}. Fig.~\ref{fig:E-Lz} shows the H3 giants (black points) and the wide binaries (red points) in the $E_{tot}$-$L_z$ space. We select H3 giants (black points) by $\log g<3.5$ and adopt the same criteria in Sec.~\ref{sec:h3} and Sec.~\ref{sec:cri-kin} to ensure reliable kinematic measurements, except that we include giant stars from the Sagittarius (\texttt{Sgr\_FLAG==1}). Applying our wide binary search to stars with \texttt{Sgr\_FLAG==1} results in two additional wide binaries, but both of them do not satisfy the kinematics criterion in Sec.~\ref{sec:cri-kin}, making no change to Fig.~\ref{fig:E-Lz}. The distance distributions vary in the $E_{tot}$-$L_z$ space. In particular, the thick disk stars are on average closer to the Sun than the halo stars. Using the H3 giants yields a relatively unbiased view with respect to the distances of the substructures in the halo out to $\sim50$\,kpc \citep{Naidu2020}.

Two main structures in Fig.~\ref{fig:E-Lz} are the thick disk and the {\it Gaia}-Sausage-Enceladus (the overdensity centered at $L_z=0$) remnant. There are more other substructures related to previous accretion events in the halo detailed in \citet{Naidu2020}. These substructures mark the rich accretion history of the Milky Way. In particular, {\it Gaia}-Sausage-Enceladus is from a merger event with a mass ratio of 1:4 about 10\,Gyr ago \citep{Belokurov2018,Helmi2018,Bonaca2020,Koppelman2020,Naidu2021}, accounting for $\sim50$\% of the halo population within 20\,kpc of the inner Milky Way \citep{Naidu2020}. The merger of {\it Gaia}-Sausage-Enceladus likely dynamically heated up the Milky Way's disk and formed the thick disk \citep{Gallart2019}. Then the Sagittarius dwarf galaxy started infalling around 8\,Gyr ago with a mass ratio of $\sim1$\% \citep{Ibata1994, Law2010, Dierickx2017,Fardal2019}. Since then, the Milky Way has undergone a relatively quiescent evolution and grown the low-$\alpha$ (thin) disk \citep{Wyse2001,Wyse2009,Ting2019a,Bonaca2020}.

In Fig.~\ref{fig:E-Lz}, we show all \Nwbgoodkinematics\ wide binaries that pass the kinematics quality criteria. We use the $E_{tot}$-$L_z$ plane to separate the halo and the thick-disk populations. Due to H3's target selection, 90\% of the H3 stars have Galactic heights $|Z|>1$\,kpc, and therefore we refer to the disk stars in the H3 as thick-disk stars. The solid blue line in Fig.~\ref{fig:E-Lz} is our empirical demarcation line, with thick disk stars located in the lower-left part and the halo stars for the rest. The disk-star selection is: (1) $E_{tot}<-0.7\times10^5$\,km$^2$\,s$^{-2}$; $\land$ (2) $L_{z}<0$; $\land$ (3) $E_{tot, 5}<-1.45 - 0.22 L_{z,3}$, where $E_{tot, 5}=E_{tot} / $($10^5$\,km$^2$\,s$^{-2}$) $L_{z, 3} = L_z /$($10^3$\,kpc\,km\,s$^{-1}$), and $\land$ is the Boolean `and' operator. With this selection, we have \Nwbdisk\ (96\%) thick-disk wide binaries and \Nwbhalo\ (4\%) halo wide binaries. If we apply this disk-halo selection to the entire H3 catalog with the same quality criteria, 16\% of the H3 stars are classified as halo stars. One main reason for the lower efficiency of identifying wide binaries in the halo than in the disk is that halo stars are more distant, making {\it Gaia} more difficult to resolve or to detect the companion. The other reason is their different wide binary fractions, which will be detailed in the next Section.

\begin{figure}
	\centering
	\includegraphics[width=1\linewidth]{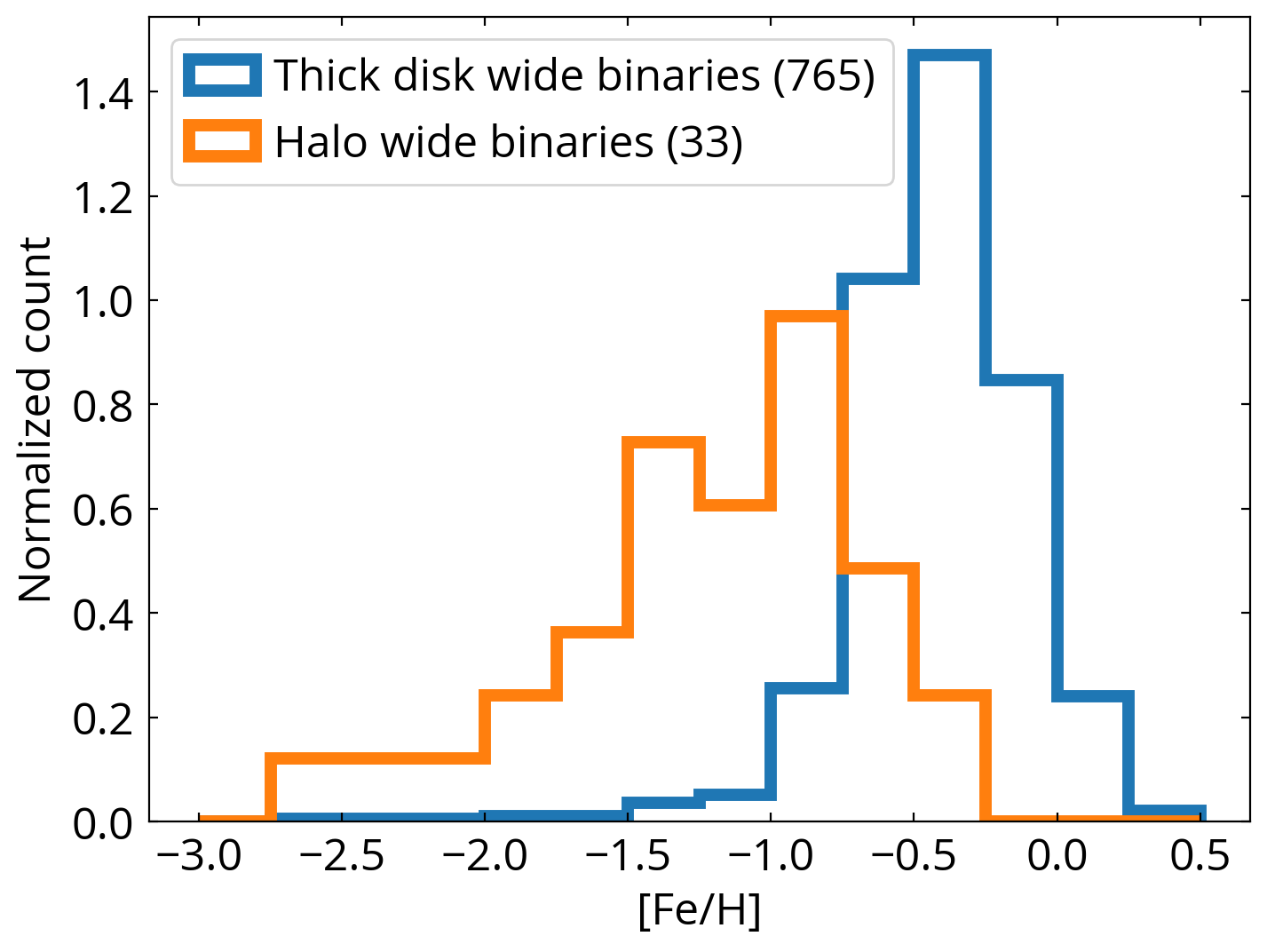}
	\caption{The metallicity distribution of the H3 wide binaries. The most metal-poor wide binary in our sample has [Fe/H]$=-2.7$. The [Fe/H] of the halo wide binaries is consistent with that of {\it Gaia}-Sausage-Enceladus. }
	\label{fig:feh-wb}
\end{figure}

Fig.~\ref{fig:feh-wb} shows the normalized distribution of surface iron abundances for disk and halo wide binaries. Halo wide binaries have [Fe/H] peaking around $-1$, similar to the metallicity of {\it Gaia}-Sausage-Enceladus remnant \citep{Naidu2020}. Wide binaries in the thick disk have [Fe/H] centering around $-0.5$. Both resolved triples belong to the disk with a relatively high [Fe/H] of $-0.38$ and $+0.01$. The most metal-poor wide binaries in our sample are [Fe/H]$=-2.7$ in the disk and [Fe/H]$=-2.5$ in the halo (the distant giant twin in Fig.~\ref{fig:HR}), among the most metal-poor wide binaries known today \citep{Reggiani2018,Hwang2021a,Lim2021}.

It is particularly interesting that many halo wide binaries have $E_{tot}$ and $L_z$ associated with the {\it Gaia}-Sausage-Enceladus remnant. The stars from the {\it Gaia}-Sausage-Enceladus are characterized by their high Galactic eccentricities $e>0.7$ \citep{Belokurov2018,Naidu2020}. Indeed, 26 out of the 33 halo wide binaries have $e>0.7$, suggesting that they are from the {\it Gaia}-Sausage-Enceladus remnant. The rest of 7 low-eccentricity ($e<0.7$) halo wide binaries may belong to the in-situ halo, and 3 of them have [Fe/H] and [$\alpha$/Fe] satisfying the criteria of the in-situ halo in \cite{Naidu2020}. Together with a recent finding of a wide binary associated with the Sequoia remnant \citep{Lim2021}, our results unambiguously show the existence of wide binaries among the accreted stars.

The two halo wide binaries having the largest $E_{tot}$ in Fig.~\ref{fig:E-Lz} are the ones with distances $>10$\,kpc discussed earlier in Sec.~\ref{sec:dist-separation} and ~\ref{sec:HR}. The one with the largest $E_{tot}$ has a highly eccentric Galactic orbit ($e=0.92$) and may be considered as {\it Gaia}-Sausage-Enceladus' member in \cite{Naidu2020}, despite of its unusually high $E_{tot}$. The other one has a modest Galactic eccentricity of 0.47 does not belong to the {\it Gaia}-Sausage-Enceladus. Its kinematics is similar to the Sagittarius in $E_{tot}$-$L_z$ space, but its $L_y$ and $L_z$ do not fall in the selection for Sagittarius stars in \cite{Johnson2020}.

There are three metal-poor wide binaries with [Fe/H]$<-2$ in the disk. All of them are main-sequence stars with ages $\sim10$\,Gyr according to the \texttt{MINESweeper} spectroscopic-photometric estimates, high [$\alpha$/Fe] (0.56, 0.54, 0.37), and prograde orbits ($L_z<0$). Although the sample is limited, they may be related to the ancient Milky Way disk \citep{Carter2021}.

\section{The abundance dependence of the wide binary fraction}
\label{sec:wbf-abundance}

\subsection{Bayesian formulation}
\label{sec:baye}

Our purposes are twofold: we aim to (1) compare the halo and disk wide binary fraction at a fixed metallicity, and to (2) investigate the metallicity trend of the disk wide binary fraction. Due to the small sample size and the limited metallicity range, we are not able to investigate the metallicity dependence of the wide binary fraction within the halo sample. These investigations need to take distances and masses into account. At a fixed binary separation, it is more difficult to spatially resolve more distant binaries, and companions of more distant binaries are more likely to fall below the magnitude limit of {\it Gaia}. Furthermore, because H3 selects targets based on apparent magnitudes, different metallicity and mass ranges are probed at different distances. Therefore, to compare wide binary fractions at different metallicities, we need to include the distance and mass dependence in the model.

For an H3 star $i$, we model the probability that this star has a wide binary companion detected by {\it Gaia} by
\begin{equation}
\label{eq:p-single}
\begin{split}
    & p_i(w_i=1|\bm{\theta}) \\
    & = f_{wb}({\rm [Fe/H]}_i, {\rm [\alpha/Fe]}_i)\times m_i^{a} \times d_i^{b},
\end{split}
\end{equation}
where $w_i$ is a Boolean value indicating if the star $i$ is a wide binary ($w_i=1$) or not ($w_i=0$) in our sample, and $m_i$ and $d_i$ are the mass and the distance of the star $i$ in units of solar mass and kpc. The probability that a star $i$ does not have a resolved {\it Gaia} companion is $p_i(w_i=0|\bm{\theta})=1-p_i(w_i=1|\bm{\theta})$. $\bm{\theta}$ is the set of the model parameters including the exponents $a$ and $b$ and other parameters involved in $f_{wb}$ which will be introduced shortly. We adopt a power-law form for the mass and distance dependence, with exponents $a$ and $b$. The distance dependence comes from the selection efficiency due to the angular resolution and the magnitude limit, and their relation depends on the underlying binary separation distribution. We run a simulation for wide binaries with companions following the Kroupa initial mass function \citep{Kroupa2001IMF} from 0.2 to 1\Msun\ and a power-law binary separation distribution $n(s)\propto s^{-2}$ from $s=10^2$ to $10^5$\,AU. The mass is then converted to the absolute $G$-band magnitude using the \texttt{MIST} isochrone with an assumption that they are main-sequence stars. Adopting an angular resolution of 0.5\arcsec\ and a $G$-band magnitude limit of 20\,mag, the simulation finds that a power-law form is a robust description ($b\sim-1.3$ for this particular case) for the distance dependence of the detection fraction at distances between 0.5 to 5\,kpc. 

$f_{wb}({\rm [Fe/H]}_i, {\rm [\alpha/Fe]}_i)$ in Eq.~\ref{eq:p-single} is the wide binary fraction at different chemical abundances for star $i$. We assume that the [Fe/H] and [$\alpha$/Fe] dependences are independent and therefore the function is separable:
\begin{equation}
\label{eq:p-feh}
    f_{wb}({\rm [Fe/H]}_i, {\rm [\alpha/Fe]}_i) = g({\rm [Fe/H]}_i) \times h({\rm [\alpha/Fe]}_i).
\end{equation}

To investigate the different iron abundance dependence between halo and disk stars, we define $g({\rm [Fe/H]}_i)$ as 
\begin{equation}
\label{eq:g-halo-disk}
    g({\rm [Fe/H]}_i) =\begin{cases}
    g_{halo}, & \text{if halo star}.\\
    g_{disk}({\rm [Fe/H]}_i), & \text{if disk star}.
  \end{cases}
\end{equation}
We use a single value $g_{halo}$ for the halo sample because of its small sample size and limited abundance range. For disk stars, we consider a multi-step function such that $g_{disk}({\rm [Fe/H]}_i)=g_{disk,j}$ if $b_j<{\rm [Fe/H]}_i<b_{j+1}$. Based on the sample size, the metallicity bin $b_j$ is chosen to be [$-3$, $-2$, $-1$, $-0.5$, 0, 0.5]. In this way, we can measure the disk wide binary fraction in each metallicity bin.

The dependence on the $\alpha$-captured abundance is encoded in $h({\rm [\alpha/Fe]}_i)$in Eq.~\ref{eq:p-feh} by
\begin{equation}
\label{eq:p-afe}
    h({\rm [\alpha/Fe]}_i) = 1 + c \times ({\rm [\alpha/Fe]}_i - 0.25),
\end{equation}
where $c$ is the only free parameter for the slope of the dependence on [$\alpha$/Fe], and $c=0$ means no dependence on [$\alpha$/Fe]. Because [Fe/H] is the main focus here, we choose a simpler form for [$\alpha$/Fe] than [Fe/H] to reduce the number of free parameters involved, and we use the same [$\alpha$/Fe] relation for both disk and halo stars. 

Therefore, $\bm{\theta}$ in Eq.~\ref{eq:p-single} includes 9 free parameters: $a$ and $b$ for the mass and distance dependence, $c$ for the [$\alpha$/Fe] dependence (Eq.~\ref{eq:p-afe}), one from $g_{halo}$, and 5 from $g_{disk,j}$. We do not marginalize the uncertainties of stellar parameters ([Fe/H]$_i$, etc.) because their uncertainties from H3 are small and do not significantly affect the results.

$f_{wb}$ in Eq.~\ref{eq:p-single} can be understood as the observed fraction of binaries with separations of about $10^3$-$10^4$\,AU in the H3 data anchored at 1\Msun\ stars at a distance of 1\,kpc from the Sun. In other words, for a 1\Msun\ star ($m_i$=1) at a distance of 1\,kpc ($d_i$=1), the probability of this star having an observed wide companion is $p_i=f_{wb}({\rm [Fe/H]}_i, {\rm [\alpha/Fe]}_i)$, depending on whether it is a disk or halo star and its iron and $\alpha$-captured abundances (Eq.~\ref{eq:p-feh}-\ref{eq:p-afe}). For stars with arbitrary masses ($m_i$) and distances ($d_i$), the probability of having an observed companion ($p_i$) depends on the wide binary fraction $f_{wb}({\rm [Fe/H]}_i, {\rm [\alpha/Fe]}_i)$ and the distance and mass dependence controlled by the power-law indices $a$ and $b$ (Eq.~\ref{eq:p-single}).  

The Bayesian formulation is 
\begin{equation}
\begin{split}
\label{eq:p-baye}
    p(\bm{\theta}|\{w_i\}) \
    \propto p(\bm{\theta}) \prod_i p_i(w_i|\bm{\theta}),
\end{split}
\end{equation}
where $p(\bm{\theta}|\{w_i\})$ is the posterior distribution of the parameter set $\bm{\theta}$ given $\{w_i\}$, the observed wide binary identification for the H3 sample. We use uninformative flat priors for $p(\bm{\theta})$ and exclude the parameter space that makes the probability $p_i>1$ for any $i$ in Eq.~\ref{eq:p-single}. Eq.~\ref{eq:p-baye} can be rewritten as 
\begin{equation}
\begin{split}
\label{eq:p-baye2}
    p(\bm{\theta}|\{w_i\}) \
    \propto \ 
    \prod_{w_i==1}  p_i(w_i=1|\bm{\theta}) \ 
    \prod_{w_i==0}  \left [1-p_i(w_i=1|\bm{\theta})\right],
\end{split}
\end{equation} 
where $p_i(w_i=1|\bm{\theta})$ is from Eq.~\ref{eq:p-single}. We use \texttt{emcee} \citep{Foreman-Mackey2013}, an affine-invariant ensemble sampler for Markov chain Monte Carlo, to sample the posteriors of Eq.~\ref{eq:p-baye2}. To distinguish halo and disk stars, we require the sample to have good Galactic kinematics with criteria described in Sec.~\ref{sec:cri-kin}. We limit the sample to heliocentric distances $<5$\,kpc because wide binary detectability beyond this distance is low and may not follow the power-law distance dependence we adopt. Two resolved triples are excluded from this analysis. Therefore, this analysis has 8610 halo stars (23 halo wide binaries) and 80572 disk stars (757 disk wide binaries). The resulting corner plot (the two-dimensional posterior distribution) made using \texttt{corner} \citep{Foreman-Mackey2016} is shown in the Appendix, Fig.~\ref{fig:corner}.

Our formulation includes the mass and distance dependence of binary detectability in Eq.~\ref{eq:p-single}. This is important because halo and disk stars have different distance and mass distributions due to their physical differences and H3's target selection. This formulation assumes that the wide binary fractions of all stars follow the same mass, distance, and [$\alpha$/Fe] dependence. This implicitly assumes that the mass-ratio distribution is not a strong function of metallicities and binary separations. This mass-ratio assumption is validated by \cite{Moe2017} that the mass-ratio distribution for binaries at $>200$\,AU is consistent with random pairings from the initial mass function. In addition, we assume that the [Fe/H] and [$\alpha$/Fe] are independent and therefore are separable in Eq.~\ref{eq:p-feh}. Future studies are needed to investigate these assumptions.

\subsection{Wide binary fraction in the halo and the disk, and its metallicity dependence}
\label{sec:feh-dep}

\begin{figure*}
	\centering
	\includegraphics[width=0.8\linewidth]{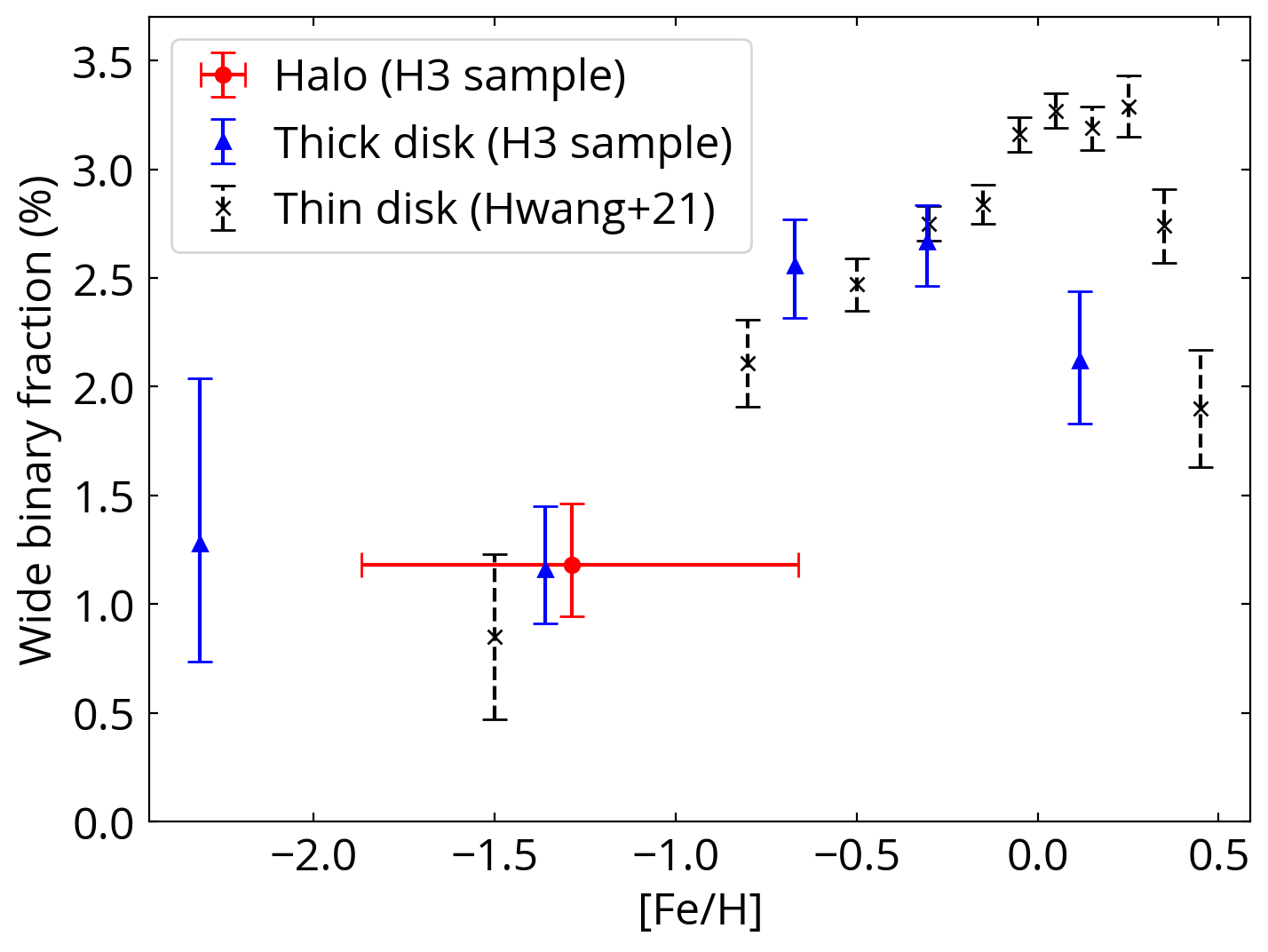}
	\caption{The wide binary fraction as a function of [Fe/H]. The blue and red markers show the results from this study using the H3 survey, with [Fe/H] values indicating the mean metallicity in each sample. The red point is for the halo stars, and its horizontal error bar indicates the 16-84 percentiles of its metallicity distribution. The blue triangles are for the thick-disk stars. The wide binary fractions of the halo (red) and the thick disk (blue) can be compared in the absolute sense because they are derived from the same data with the same model. The black crosses with dashed error bars show the results from a thin-disk-dominated sample \citep{Hwang2021a}. Because the thin-disk results are derived from a different sample, we can only compare the overall relative metallicity trend with the thick-disk and halo results, but not the absolute values of the wide binary fractions. This plot shows that the wide binary fractions of the halo and the thick disk are consistent at a fixed [Fe/H]. The different trend between the thin and thick disk at [Fe/H]$>-0.5$ is mainly due to the younger stellar population in the thin disk \citep{Hwang2021a}. } 
	\label{fig:wbf}
\end{figure*}

Fig.~\ref{fig:wbf} shows the wide binary fraction as a function of [Fe/H]. The red circle and blue triangles show the measurements for the halo and the thick disk, respectively. Their [Fe/H] values are the mean metallicities in each sample, and the horizontal error bar of the halo sample indicates the 16-84 percentiles of its metallicity distribution. Their wide binary fractions are $g_{halo}$ and $g_{disk}$([Fe/H]) in Eq.~\ref{eq:p-feh}, defined as the observed values at 1\,kpc for 1\Msun\ stars. The wide binary fractions and their errors in Fig.~\ref{fig:wbf} are the 16-50-84 percentiles of their marginalized distributions.

For comparison, in Fig.~\ref{fig:wbf} we overplot the results from \cite{Hwang2021a} as black crosses where the 500-pc sample is dominated by the thin-disk stars and the wide binaries refer to separations between $10^3$ and $10^4$\,AU, with [Fe/H] values indicating the centers of the bins. The wide binary fractions of the disk and halo stars from this work (red circle and blue triangles) can be compared with each other in the absolute sense, meaning that the halo wide binary fraction is consistent with that in the thick disk at a fixed [Fe/H], taking into account the differences in the distance and mass distributions. Although the thick-disk wide binary fractions from this work and those from \cite{Hwang2021a} happen to have similar values in Fig.~\ref{fig:wbf}, they should only be compared in a relative sense (i.e. the overall metallicity trend) but not in an absolute sense, because they use samples that have different detection limits for the companions due to the different distance, binary separation, and mass ranges.

Fig.~\ref{fig:wbf} shows that the wide binary fraction in the thick disk decreases towards the low-metallicity end. This is similar to the trend in the thin disk \citep{Hwang2021a}, but now this work extends the trend down to [Fe/H]$<-2$. At [Fe/H]$>-0.5$, the H3 sample shows a flat relation between the wide binary fraction and the metallicity, in contrast to the peak at [F/H]$=0$ for the thin-disk stars. This is consistent with \cite{Hwang2021a} where we found that the peak at [F/H]$=0$ is more prominent for younger disk stars.

\cite{Hwang2021a} differentiated the thick-disk stars and halo stars using total 3-dimensional velocities and the maximum Galactic heights of the stellar orbits. We found that at a fixed metallicity, the wide binary fraction of the halo seems consistent with that of the thick disk. However, there are only two halo wide binary candidates in \cite{Hwang2021a}, so the conclusion is still limited by small number statistics.

Here with a better Galactic kinematics selection and a larger halo wide binary sample from the H3 survey, we robustly show that the wide binary fraction of the halo stars is consistent with that of the thick-disk stars at a fixed metallicity. This result suggests that iron abundance is the main driver of the wide binary fraction, and that the thick disk and halo stars do not have significantly different wide binary fractions once the metallicity, distance, and mass are taken into account.

The [$\alpha$/Fe] dependence is an important clue about binary formation. \cite{Mazzola2020} find that the close binary ($a<$10\,AU) fraction anti-correlates with [$\alpha$/Fe]. This relationship may arise because the dust formed from $\alpha$-captured elements affects the fragmentation of the protostellar disks by changing its opacity. In contrast, \cite{Niu2021} consider the total binary fraction and suggest a tentative positive correlation between the binary fraction and [$\alpha$/Fe] at a fixed [Fe/H].

In contrast to the previous studies, here we investigate the [$\alpha$/Fe] dependence of the wide binary fraction at separations $>10^3$\,AU. The parameter for the [$\alpha$/Fe] dependence in Eq.~\ref{eq:p-afe} is measured to be $c=-0.05^{+0.06}_{-0.14}$, where the errors correspond to 16-84 percentiles.  Therefore, $c$ is consistent with zero and the wide binary fraction does not appear to depend on the $\alpha$-captured abundances. The different [$\alpha$/Fe] dependence of the close and wide binary fraction suggests that different formation mechanisms are responsible for close and wide binaries, similar to the conclusions from the [Fe/H] dependence \citep{El-Badry2019a,Hwang2021a}.

\begin{figure}
	\centering
	\includegraphics[width=1\linewidth]{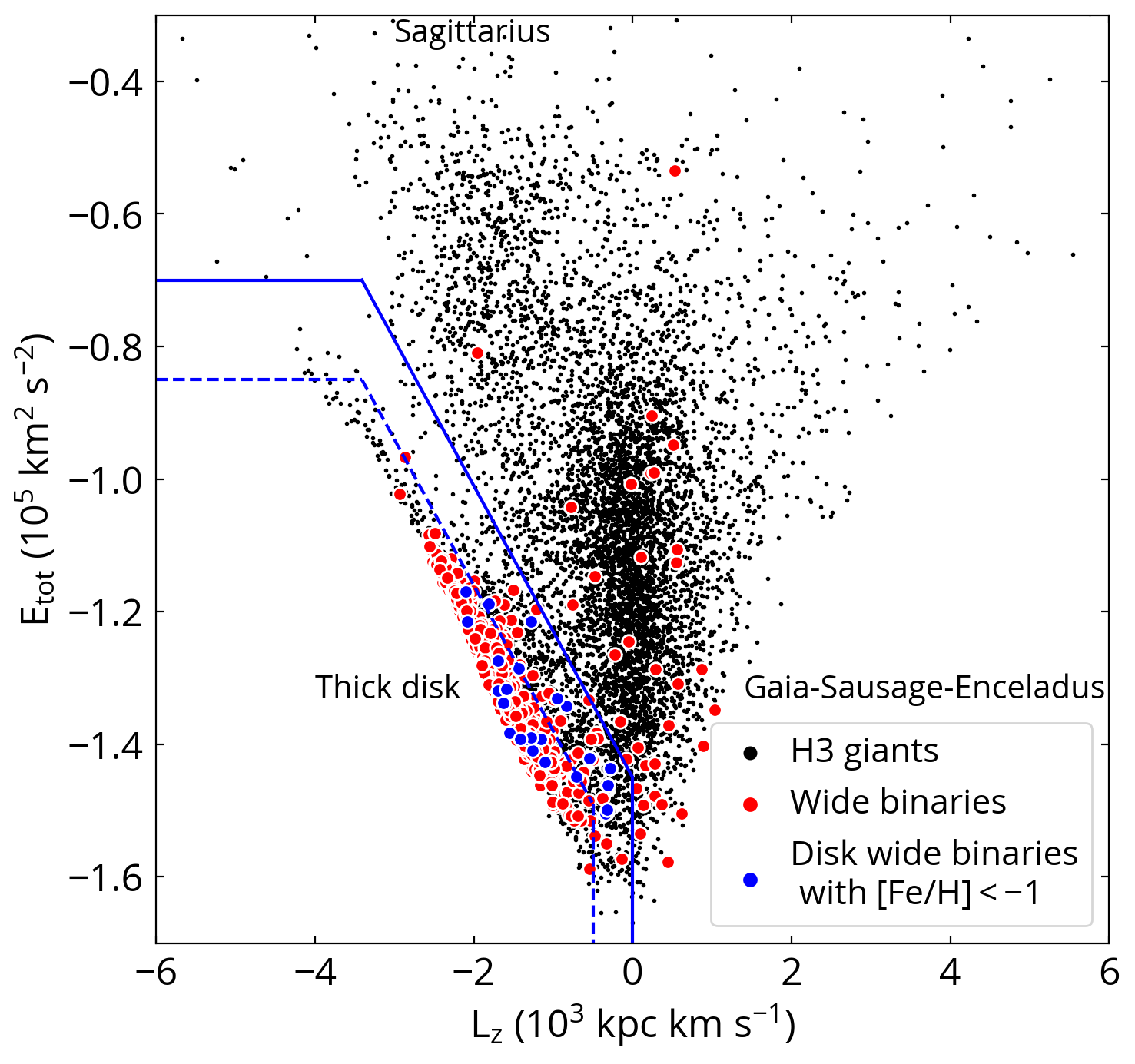}
	\caption{Same as Fig.~\ref{fig:wbf}, but with metal-poor ([Fe/H]$<-1$) disk wide binaries highlighted. We obtain similar wide binary fractions for the disk stars using a more strict disk-star selection (dashed line). Therefore, the similar wide binary fractions in the halo and the thick disk is not due to the contamination of the halo stars in the metal-poor thick-disk sample. } 
	\label{fig:E-Lz-test}
\end{figure}

One concern is whether the metal-poor disk stars at [Fe/H]$<-1$ are contaminated by halo stars, which might explain the similar wide binary fractions in the halo and the thick disk at [Fe/H]$<-1$. Fig.~\ref{fig:E-Lz-test} shows the disk wide binaries with [Fe/H]$<-1$ (blue points) in the $E_{tot}$-$L_z$ space. There is no strong tendency for metal-poor wide binaries being located close to the original demarcation line (solid blue line). While some wide binaries are closer to the demarcation line, they may still be genuine metal-poor disk stars because the metal-weak thick disk is located at a similar $E_{tot}$-$L_z$ region \citep{Naidu2020}. To test how the disk-star selection affects the result, we use a more strict disk-star selection (dashed line in Fig.~\ref{fig:E-Lz-test}): (1) $E_{tot,5}<-0.85$; $\land$ (2) $L_{z,3}<-0.5$; $\land$ (3) $E_{tot, 5}<-1.6 - 0.22 L_{z,3}$. Using this more strict disk-star selection and the same Bayesian approach from Sec.~\ref{sec:baye}, we obtain wide binary fractions of the thick disk consistent with Fig.~\ref{fig:wbf}. Therefore, the contamination of halo stars in the disk selection plays a minor role in our results. 

\section{Discussion}

\label{sec:discussion}

\subsection{Comparison with the literature}

In the pre-Gaia era when high-precision parallaxes were not available for most of the stars, the reduced proper motion diagram was used to distinguish high-velocity halo stars from lower-velocity disk stars and to identify halo wide binaries. The reduced proper motion diagram uses proper motions as a proxy for distances because closer stars have larger proper motions, and halo stars are more separated from the disk stars because of halo stars' lower metallicities and higher spatial velocities \citep{Salim2002}. Using this method, previous studies have found 100-200 halo(-like) wide binary candidates \citep{Chaname2004,Quinn2009,Allen2014,Coronado2018}. Wide binaries in the halo can be used to constrain the mass of massive compact halo objects -- for example, the contribution of free-floating black holes and neutron stars to the mass budget of the Universe \citep{Yoo2004,Quinn2009a,Monroy-Rodriguez2014}.

In addition to proper motions, the reduced proper motion diagram uses photometry to distinguish halo and disk stars. However, because lower-metallicity stars have a higher close binary fraction \citep{Moe2019}, and close binaries are more likely to have wide companions \citep{Hwang2020c}, metal-poor stars selected from the reduced proper motion diagram may have binary-related selection effects because unresolved companions make the systems deviate from the reduced proper motion diagram.

Gaia provides high-precision astrometry for billions of stars \citep{Gaia2016}, making the search for a large number of wide binaries possible \citep{Oh2017,El-Badry2018b,Jimenez-Esteban2019,Hartman2020}. Using {\it Gaia} EDR3, \cite{El-Badry2021} have cataloged about a million wide binaries within 1\,kpc. By cross-matching this catalog the with LAMOST survey DR6 \citep[][]{Deng2012, Zhao2012} and using LAMOST's radial velocities to compute $E_{tot}$ and $L_z$, we find about 50 halo wide binaries that satisfy the halo selection from Sec.~\ref{sec:kinematics} and SNR $>10$ in g-band from LAMOST with a lower spectral resolution ($R\sim1800$) than H3.

Our wide binaries feature the most distant ($>1$\,kpc and four candidates at $>10$\,kpc), most metal-poor (down to [Fe/H]$=-2.7$) sample with detailed abundance and kinematic measurements from a $R=32,000$ spectroscopy. The H3 survey is critical in providing a homogeneous sample of disk and halo stars to compare the wide binary fractions among various subsamples. Since our wide binary search depends only on astrometry and not on photometry, our selections are less affected by unresolved companions. Our halo classifications and the association with {\it Gaia}-Sausage-Enceladus remnant are robust thanks to the precise $E_{tot}$ and $L_z$ measurements from H3.

\subsection{Origin of the iron-abundance dependence}

\cite{Hwang2021a} identified 7671 wide binaries within 500\,pc with projected separations between $10^3$ and $10^4$\,AU. These wide binaries are mostly thin- and thick-disk stars. With the iron abundances measured from LAMOST \citep{Wu2011,Wu2011a,Ting2019,Xiang2019}, we found that the wide binary fraction at $10^3$-$10^4$\,AU peaks at around [Fe/H]$=0$, and decreases towards both low and high metallicity ends (black symbols in Fig.~\ref{fig:wbf}).

\citet{Hwang2021a} proposed several possibilities to explain such metallicity dependence. The positive correlation between [Fe/H] and the wide binary fraction at [Fe/H]$<0$ may be due to the denser formation environments in the earlier Universe that tend to disrupt the wide binaries. At [Fe/H]$>0$, the anti-correlation between [Fe/H] and the wide binary fraction may be due to some wide binaries forming from the dynamical unfolding of compact triples \citep{Reipurth2012,Elliott2016}, and thus they inherit the metallicity dependence from the close binary fraction \citep{Raghavan2010, Yuan2015, Badenes2018, Moe2019, El-Badry2019a}. The radial migration of Galactic orbits \citep{Sellwood2002} may cause such non-monotonic metallicity dependence, either due to different formation environments at different galactocentric radii or due to disruption of wide binaries during the enhanced interaction with high-density gas clumps during the migration process. Wide binary disruptions by passing stars are unlikely to play an important role because the disruption timescale of $\sim100$\,Gyr for $\sim1000$\,AU binaries is too long \citep{Weinberg1987}.

If the metallicity dependence of the wide binary fraction is due to the radial migration, which is more important for disk stars, we would expect a higher wide binary fraction in the halo than in the disk. Our H3 results suggest that iron abundance is the main driver for the wide binary fraction (Fig.~\ref{fig:wbf}), instead of their origin (disk versus halo) nor $\alpha$-captured abundance. The similar wide binary fractions in the halo and the disk implies that radial migration is not the main cause for the metallicity dependence of the wide binary fraction at [Fe/H]$<0$.

Our leading explanation for the iron abundance dependence of the wide binary fraction at [Fe/H]$<0$ is that the formation environments are denser at lower [Fe/H], which are more disruptive for wide binaries. In particular, studies have found that the star formation environments have higher pressures and densities at higher redshifts, and higher-mass clusters tend to form in such environments \citep{Harris1994,Elmegreen1997, Kravtsov2005, Kruijssen2014, Ma2020}. Therefore, the lower wide binary fraction may reflect their denser formation environments in the earlier epochs.

Although a lower-metallicity population is on average older than a higher-metallicity sample \citep{Casagrande2011, Bensby2014,SilvaAguirre2018}, our result is better explained by the metallicity dependence instead of the age dependence. Following \cite{Bonaca2020}, we select main-sequence turn-off stars by $3.8<\log\ g<4.3$ and SNR$>10$ for their precise age measurements. This sample shows that, for our H3 thick-disk sample, the mean stellar age changes approximately linearly from $12$\,Gyr at [Fe/H]$=-2.0$\,dex \citep{Carter2021} to $8.5$\,Gyr at [Fe/H]$=0$\,dex. This does not explain the metallicity trend in Fig.~\ref{fig:wbf} where the wide binary fraction becomes flat at [Fe/H]$>-1$ when the mean age changes by about 2\,Gyr from [Fe/H]$=-1$ to [Fe/H]$=0$. Therefore, our results suggest that stellar ages play a minor role.

While halo wide binaries are currently not susceptible to disruptions via radial migration and giant molecular cloud, they were once in presumably gas-rich environments in the progenitor (dwarf) galaxies. It is intriguing that halo and thick-disk wide binary fractions are similar at a fixed metallicity despite their potentially different formation environments. The similar wide binary fractions between the thick disk and the halo may hint that the metallicity and the formation environment properties are tightly correlated, and by controlling the metallicity, we mitigate the differences in their formation environments. One important formation environment property is the cluster mass function because wide binaries are sensitive to the density (which correlates with mass) of clusters (e.g. \citealt{Parker2009,Geller2019}). Therefore, one possibility is that the Milky Way and the {\it Gaia}-Sausage-Enceladus dwarf may have similar cluster mass functions at a fixed metallicity (\citealt{Krumholz2019}), resulting in the similar wide binary fractions.

Since wide binaries are unlikely to survive in globular clusters (e.g. \citealt{Ivanova2005}), our results can place a rough constraint on the contribution of disrupted globular clusters to the stellar halo. Here we find that the wide binary fraction of the halo is consistent with that of the thick disk (Fig.~\ref{fig:wbf}). Assuming that the original halo wide binary fraction (halo stars that are not from disrupted globular clusters) is the same as that in the disk at a fixed metallicity and that the wide binary fraction from disrupted globular clusters is zero, our result implies that the disrupted globular clusters contribute less than $\sim50$\% of the halo. If the disrupted globular clusters contribute more than $50$\% of the halo, the halo wide binary fraction would be significantly lower than that of the thick disk, inconsistent with our results. The $<50$\% constraint is consistent with other studies finding that $\sim10$\% of halo stars originate in disrupted globular clusters \citep{Martell2010,Martell2011,Martell2016,Koch2019}, or likely an even smaller fraction \citep{Naidu2020}. Our constraint serves as a useful independent check but is likely tentative because of the various assumptions used, and it may be possible that wide binaries can form during globular cluster disruption \citep{Penarrubia2021}.

\section{Conclusions}
\label{sec:conclusion}

What determines the wide binary fraction remains not well understood. \citet{Hwang2021a} found that the wide binary fraction of disk stars peaks around solar metallicity ([Fe/H]$=0$) and decreases toward both low and high metallicity end. We proposed three possible scenarios that may explain this metallicity dependence, including the radial migration of Galactic orbits. In this paper, armed with the thick-disk and halo wide binaries identified from the H3 survey, we rule out the radial migration scenario at [Fe/H]$<0$. Our findings include:

\begin{enumerate}
	\item With {\it Gaia}'s astrometry, we search for resolved wide companions around the H3 stars (Fig.~\ref{fig:WBsearch-one}). Our search results in a total of \Nwidebinary\ wide binaries and two resolved triples. The photometry of the non-H3 companions are consistent with the isochrones of the H3 stars, suggesting that most of them are genuine wide binaries instead of chance-alignment pairs (Fig.~\ref{fig:dbprp}).
	\item Most of the wide binaries are main-sequence stars (Fig.~\ref{fig:HR}), with binary separations ranging from $10^3$ to $10^5$\,AU and heliocentric distances $\gtrsim1$\,kpc (Fig.~\ref{fig:sep-dist}). These wide binaries have [Fe/H] down to $-2.7$ (Fig.~\ref{fig:feh-wb}), among the most metal-poor wide binary currently known. There are four distant wide binary candidates at $>10$\,kpc. They are all rare double-giant wide binaries, and one of them is an unusual metal-poor ([Fe/H]$=-2.5$) giant twin where the member stars have very similar photometry (Fig.~\ref{fig:HR}).
	\item Based on the Galactic kinematics, we classify the wide binaries into disk and halo population (Fig.~\ref{fig:E-Lz}), resulting in a sample of \Nhalowb\ halo wide binaries. The majority of these halo wide binaries have kinematics associated with the accreted {\it Gaia}-Sausage-Enceladus remnant. 
	\item The wide binary fraction in the disk decreases toward the low-[Fe/H] end (Fig.~\ref{fig:wbf}), consistent with the result from \cite{Hwang2021a}. We do not find a significant dependence of the wide binary fraction on [$\alpha$/Fe]. Both [Fe/H] and [$\alpha$/Fe] dependences of the wide binary fraction are different from those of the close binary fraction, suggesting that different formation mechanisms are responsible for close and wide binaries. 
	\item We show that the wide binary fraction in the halo is consistent with that in the thick disk at a fixed [Fe/H] (Fig.~\ref{fig:wbf}). Since halo stars are not subject to radial migration, the similar wide binary fractions in the thick disk and the halo imply that radial migration of Galactic orbits plays a minor role in the metallicity dependence of the wide binary fraction at [Fe/H]$<0$. Our study favors the explanation that lower-metallicity formation environments have higher stellar densities that tend to disrupt wide binaries, thus lowering the wide binary fraction.

\end{enumerate}

The authors are grateful to the referee for the constructive comments. HCH appreciates the inspiring discussion with Yao-Yuan Mao on the dark matter subhalos, and with Amina Helmi on exploring other Galactic orbital parameters. HCH acknowledges support from Space@Hopkins and the Infosys Membership at the Institute for Advanced Study. YST acknowledges financial support from the Australian Research Council through DECRA Fellowship DE220101520. The H3 Survey acknowledges funding from NSF grant AST-2107253.

{\it Facilities:} Gaia, MMT.

{\it Software:} \texttt{IPython} \citep{ipython2007}, \texttt{jupyter} \citep{jupyter2016}, \texttt{Astropy} \citep{Astropy2013,Astropy2018}, \texttt{numpy} \citep{numpy2020}, \texttt{scipy} \citep{scipy2020}, \texttt{emcee} \citep{Foreman-Mackey2013,Foreman-Mackey2016},  \texttt{matplotlib} \citep{matplotlib2007}.

\section*{Data Availability}
The data underlying this article are available in the article and in its online supplementary material.

\bibliography{wide_binary_H3}{}
\bibliographystyle{aasjournal}

\appendix

\restartappendixnumbering
\setcounter{figure}{0}

\section{The posterior distributions for the binary fraction model}
\label{sec:corner}
\begin{figure}
	\centering
	\includegraphics[width=0.9\linewidth]{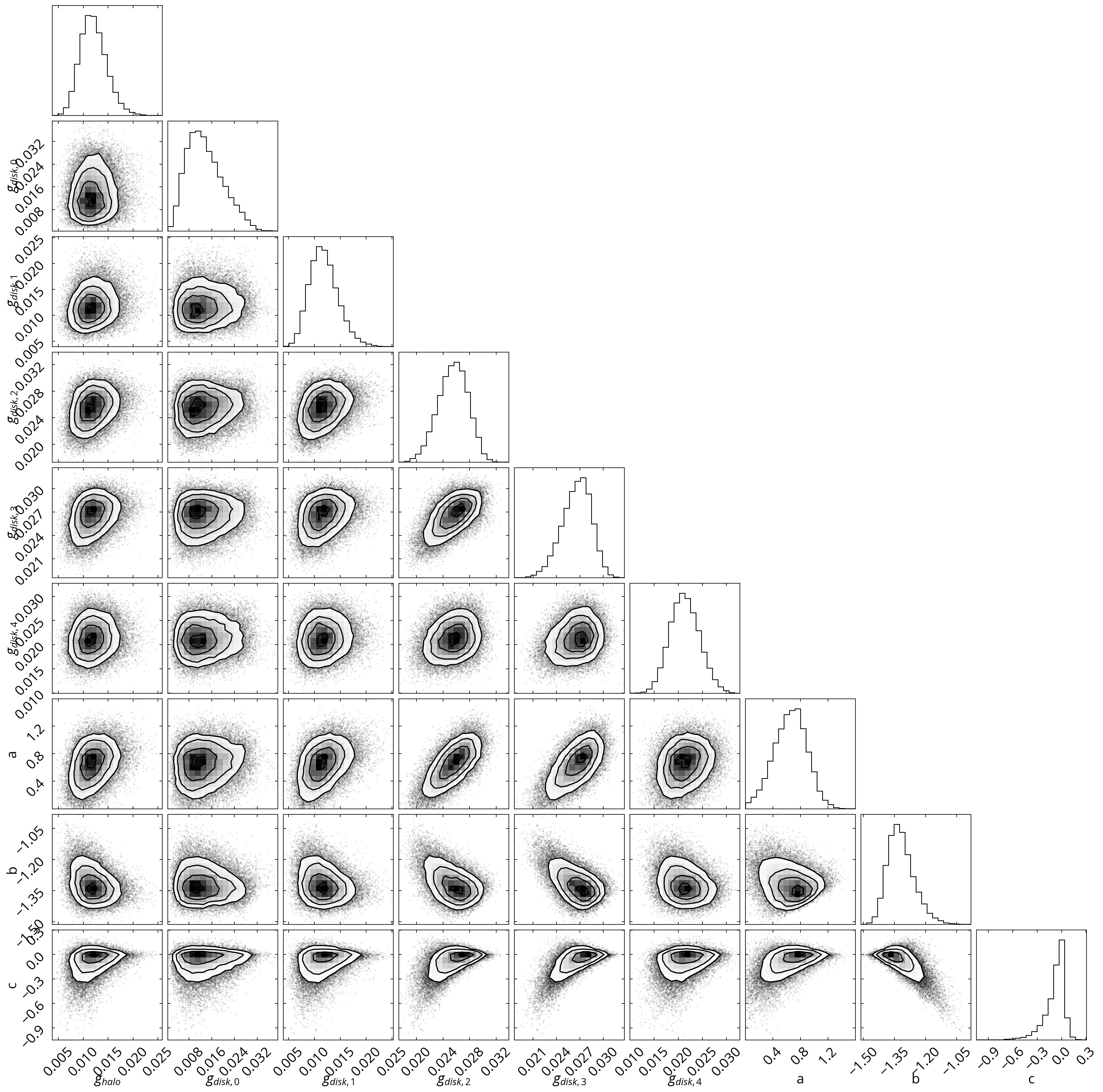}
	\caption{Corner plot for the posterior distributions of the wide binary fraction model in Sec.~\ref{sec:baye}.  }
	\label{fig:corner}
\end{figure}

Fig.~\ref{fig:corner} shows the posterior distributions of the model parameters described in Sec.~\ref{sec:baye}. Their measurements (marginalized 16-50-84 percentiles) are: $g_{halo}$ $=0.012^{+0.003}_{-0.002}$, $g_{disk,0}$ $=0.013^{+0.008}_{-0.005}$, $g_{disk,1}$ $=0.012^{+0.003}_{-0.002}$, $g_{disk,2}$ $=0.026^{+0.002}_{-0.002}$, $g_{disk,3}$ $=0.027^{+0.002}_{-0.002}$, $g_{disk,4}$ $=0.021^{+0.003}_{-0.003}$, a $=0.661^{+0.225}_{-0.252}$, b $=-1.326^{+0.070}_{-0.055}$, c $=-0.052^{+0.063}_{-0.142}$.

As a sanity check for the fitting result, for disk stars with $-0.5<$[Fe/H]$<0$ and distances between 0.8 and 1.2\,kpc, the observed wide binary fraction is $101/4279=2.36 \pm 0.23$\%. This sample has a mean mass of $m=0.77$\Msun\ and a mean distance of $d=1.05$\,kpc. The wide binary fraction from our fitting model is $g_{disk,3}\times m^a \times d^b=0.027\times 0.77^{0.661} \times 1.06^{-1.326} = 2.1\pm0.2$\%, consistent with the observed value. For more distant disk stars with $-0.5<$[Fe/H]$<0$ and distances between 1.5 and 2\,kpc, the observed wide binary fraction is $106/11141=0.95 \pm 0.09$\%. With a mean mass of 0.84\Msun\ and a mean distance of 1.76\,kpc, the wide binary fraction from the model is $g_{disk,3}\times m^a \times d^b=0.027\times 0.84^{0.661} \times 1.76^{-1.326} = 1.1\pm0.1$\%, again consistent with the observed value.

\end{document}